\documentclass[aps,prb,twocolumn,showpacs,amsmath,amssymb]{revtex4}
\usepackage{epsfig}
\usepackage{graphics}
\usepackage{psfrag}
\usepackage{bm}

\newcommand{\Brace}[2]{\genfrac{\{}{\}}{0pt}{}{#1}{#2}}

\def\a{\alpha}
\def\b{\beta}
\def\g{\gamma}
\def\d{\delta}

\def \e{\varepsilon}
\def\erg {\tau_{\rm erg}}
\def\h{\hbar}
\def \oo {\omega}

\def\Tr{\mbox{tr}\,}
\def\P{{\cal P}}
\def\S{{\cal S}}
\def\U{{\cal U}}
\def\l{\lambda}
\def\m{\mu}
\def\DD{\partial}

\def\dwell{\tau_{\rm d}}
\def \Diff {{\cal D}}
\def\Coop {{\cal C}}
\def\G {{\cal G}}
\def \Ga {{\cal G}_a}
\def \Gs {{\cal G}_s}

\def\A{{\mathcal A}}

\def\la {\langle}
\def\ra {\rangle}
\def \etal{{\it et~al.}}

\def \Thou{E_{\rm Th}}
\def\RC {\tau_{\rm RC}}

 \psfrag{f}{$f$} \psfrag{r}{$r$}
\psfrag{V0}{$V_0$} \psfrag{Vtilde}{$\tilde V$} \psfrag{eta}{$\eta$}
\psfrag{N1}[][][0.8]{$N_1$}\psfrag{N2}[][][0.8]{$N_2$}\psfrag{C0}[][][0.8]{$C_{g0}$}\psfrag{C1}[][][0.8]{$C_1$}
\psfrag{C2}[][][0.8]{$C_2$}
\psfrag{Fa}{$\Phi_a$}\psfrag{Fd}{$\Phi_d$}\psfrag{F}{$\Phi$}\psfrag{otimes}{$\otimes$}\psfrag{S}{${\cal
S}_{N\times N}$}\psfrag{U}[][][1]{${\cal U}_{M\times
M}$}\psfrag{NxN}{$ $}\psfrag{MxM}{$ $}\psfrag{G}{$\cal
G$}\psfrag{ho}{$\hbar\oo$}\psfrag{Cparas}{$C_{\rm
stray}$}\psfrag{Vp}[][][0.8]{$V_{\rm p}$}\psfrag{Cpump}{$C_{\rm p}$}
\begin{document}

\title{Rectification and nonlinear transport in chaotic dots and rings}
\date{\today}
\author{M.~L.~Polianski}
\email{polian@physics.unige.ch}\author{M.~B\"uttiker}
\affiliation{D\'epartement de Physique Th\'eorique, Universit\'e de
Gen\`eve, CH-1211 Gen\`eve 4, Switzerland}
\begin{abstract}
We investigate the nonlinear current-voltage characteristic of
mesoscopic conductors and the current generated through
rectification of an alternating external bias. To leading order in
applied voltages both the nonlinear and the rectified current are
quadratic. This current response can be described in terms of second
order conductance coefficients and for a generic mesoscopic
conductor they fluctuate randomly from sample to sample. Due to
Coulomb interactions the symmetry of transport under magnetic field
inversion is broken in a two-terminal setup. Therefore, we consider
both the symmetric and antisymmetric nonlinear conductances
separately. We treat interactions self-consistently taking into
account nearby gates.

The nonlinear current is determined by different combinations of
second order conductances depending on the way external voltages are
varied away from an equilibrium reference point (bias mode). We
discuss the role of the bias mode and circuit asymmetry in recent
experiments. In a photovoltaic experiment the alternating
perturbations are rectified, and the fluctuations of the nonlinear
conductance are shown to decrease with frequency. Their asymptotical
behavior strongly depends on the bias mode and in general the
antisymmetric conductance is suppressed stronger then the symmetric
conductance.

We next investigate nonlinear transport and rectification in chaotic
rings. To this extent we develop a model which combines a chaotic
quantum dot and a ballistic arm to enclose an Aharonov-Bohm flux. In
the linear two-probe conductance the phase of the Aharonov-Bohm
oscillation is pinned while in nonlinear transport phase rigidity is
lost. We discuss the shape of the mesoscopic distribution of the
phase and determine the phase fluctuations.
\end{abstract}\pacs{73.23.-b, 73.21.La,73.40.Ei, 73.50.Fq}

\maketitle
\section{Introduction}\label{sec:intro}
 A large part of modern physics is devoted to nonlinear
classical and quantum phenomena in various systems. Such effects as
the generation of the second harmonic or optical rectification are
known from classical physics, while quantum electron pumping through
a small sample due to interference of wave functions is a quantum
nonlinear effect. Experiments on nonlinear electrical transport
often combine classical and quantum contributions. A macroscopic
sample without inversion center \cite{UFN} exhibits a
current-voltage characteristic which with increasing voltage departs
from linearity due to terms proportional to the square of the
applied voltage. If now an oscillating (AC) voltage is applied, a
zero-frequency current (DC) is generated.

If the sample is sufficiently small, quantum effects can appear due
to the wave nature of electrons. The uncontrollable distribution of
impurities or small variations in the shape of the sample result in
quantum contributions to the DC which are random. For a mesoscopic
conductor with terminals $\a ,\b, ... $ we can describe the
quadratic current response in terms of second order conductances
$\G_{\a\b\g}$. They relate voltages $V_{\b,\oo}$ applied at contacts
or neighboring gates $\b$ at frequency $\oo$ to the current at zero
frequency at contact $\a$,
\begin{eqnarray}\label{eq:IV}
I_\a &=&\sum_{\b\g} \G_{\a\b\g} |V_{\b,\oo}-V_{\g,\oo}|^2.
\end{eqnarray}
The second order conductances include in
detail the role of the shape and the nearby conductors (gates).
They depend on external parameters like
the frequency of the perturbation, temperature, magnetic field  or
the connection of the sample to the environment.

We concentrate here on the quantum properties of nonlinear
conductance through coherent chaotic samples. Chaos could result
from the presence of impurities (disorder) or random scattering at
the boundaries (ballistic billiard). Due to electronic interference
the sign of this effect is generically random even for samples of
macroscopically similar shape. \cite{WW,AK,KL} When averaged over an
ensemble, the second order conductances vanish. As a consequence,
for a fully chaotic sample there is no classical contribution to the
DC and the nonlinear response is the result of the sample-specific
quantum fluctuations.

Interestingly enough, from a fundamental point of view these
fluctuations of nonlinear conductance are sensitive to the presence
of Coulomb interactions and magnetic field. While interactions
strongly affect the fluctuations' amplitude, their sign is easily
changed by a small variation of magnetic flux $\Phi$, similarly to
universal conductance fluctuations (UCF) in linear transport. More
importantly, without interactions the current (\ref{eq:IV}) through
a two-terminal sample is a symmetric function of magnetic field,
just like linear conductance. However, the idea that Coulomb
interactions are responsible for
  magnetic-field asymmetry in nonlinear current was recently
  proposed theoretically \cite{SB,SZ} and demonstrated experimentally
  in different mesoscopic systems. \cite{wei,Zumbuhl,marlow,ensslin,Bouchiat,Bouchiat_preprint}
  (Various aspects of nonlinear quantum \cite{PB,Coulomb,Tsvelik,PhysicaE} and
  classical \cite{AG} charge and spin transport \cite{Feldman} have been discussed later on.)
  It is useful to
  consider (anti) symmetric second order conductance $\Ga,\Gs$ defined as
  \begin{eqnarray}\label{eq:IVfield}
\Brace{{\mathcal G}_{s}(\Phi)}{{\mathcal G}_{a}(\Phi)} &=
&\frac{h}{\nu_s e^3}\frac{\DD^2}{2\DD \tilde
V^2}\left(\frac{I(\Phi)\pm I(-\Phi)}{2}\right)_{\tilde V\to 0},
\end{eqnarray}
where $\tilde V$ is a combination of voltages at the gates and
contacts varied in the experiment and $\nu_s$ accounts for the spin
degeneracy. We emphasize that, depending on the way voltages are
varied, experiments probe different linear combinations of second
order conductance elements $\G_{\a\b\g}$ of Eq. (\ref{eq:IV}). From
now on we will simply call $\Gs ,\Ga$ conductances and if no
confusion is possible leave out the expression "second order".

In the presence of a DC perturbation the mesoscopic averages of
antisymmetric \cite{SB,SZ} and symmetric \cite{PB,PhysicaE}
conductances vanish, and it is their sample-to-sample fluctuations
that are measured. Experiments are usually performed for strongly
interacting samples and the magnetic-field components $\Gs,\Ga$
allow one to evaluate the strength of interactions.
\cite{Zumbuhl,Bouchiat} In previous theoretical works on nonlinear
transport through chaotic dots several important issues have been
discussed using Random Matrix Theory (RMT). \cite{SB,PB,PhysicaE}
S\'anchez and B\"uttiker \cite{SB} found the fluctuations of $\Ga$
in a dot with arbitrary interaction strength at zero temperature and
broken time-reversal symmetry due to magnetic field. Polianski and
B\"uttiker considered the statistics of both $\Ga$ and $\Gs$ for
arbitrary flux $\Phi$, the temperature $T$, and the dephasing rate.
\cite{PB} The fluctuations of relative asymmetry $\A=\Ga/\Gs$ and
the role of the contact asymmetry on this quantity were discussed in
Ref.\,\onlinecite{PhysicaE}. The results of RMT approach were
compared with experimental data of Zumb\"uhl \etal\,\cite{Zumbuhl}
and Angers \etal\,\cite{Bouchiat}

Previously we considered statistics of $\Ga,\Gs$ for the dots where
only one DC voltage was varied. However, to avoid parasitic circuit
effects some experiments are performed varying several voltages
simultaneously. Surprisingly, the importance of the chosen
combination of varied voltages (bias mode) was not addressed before
in the literature. It turns out that an experiment where only one of
the voltages is varied \cite{marlow,Lofgren,Bouchiat} or two
voltages are asymmetrically shifted \cite{Zumbuhl,ensslin} measure
different combinations of nonlinear conductances $\G_{\a\b\g}$. For
example, in a weakly interacting dot in the first mode we found that
$\Gs\gg \Ga$, \cite{PhysicaE} but in the second bias mode the
fluctuations of nonlinear current are strongly reduced, so that
$\Gs\sim \Ga$.

It is also important to generalize the previous treatment of the
nonlinear current to mesoscopic systems biased by an AC-voltage at
{\it finite} frequency. The resulting DC is sometimes called
"photovoltaic current". We expect that in such mesoscopic AC/DC
converters the interactions lead to significant magnetic
field-asymmetry in the DC-signal. The rectification effect of
mesoscopic diffusive metallic microjunctions was theoretically
considered by Falko and Khmelnitskii \cite{FK} assuming that
electrons do not interact. Therefore, a magnetic-field asymmetry was
not predicted and was also not observed in subsequent experiments.
\cite{Bykov,BykovAB,Bartolo,Lin,Liu} The fact that the interactions
induce a magnetic field-asymmetry of the photovoltaic current when
the size of the sample is strongly reduced was recently demonstrated
in Aharonov-Bohm rings by Angers \etal \cite{Bouchiat_preprint}

However, it turns out that for an AC perturbation another quantum
interference phenomenon, also quadratic in voltage, random in sign
and magnetic field-asymmetric, contributes to the DC. Due to {\it
internal} AC- perturbations of the sample, the energy levels are
randomly shifted and a phenomenon commonly referred to as "quantum
pumping" \cite{pump,SAA} appears. Brouwer demonstrated that two
voltages applied out of phase
 generate pumped current linear in frequency, while a single voltage pumps
current quadratic in frequency $\oo$. \cite{pump} Although theory
usually considers small (adiabatic) frequencies, a photovoltaic
current could be induced by voltages applied at arbitrary frequency.
At small $\oo$ the pumping contribution vanishes and only the
rectification effect survives. In contrast, it is not clear what the
ratio of pumping current to rectification current is at large $\oo$.
To distinguish between different mechanisms it is therefore
important to consider rectification in a wide range of frequencies
in detail.

\begin{figure}{\psfrag{V1}{$V_1(\oo)$} \psfrag{V2}{$V_2(\oo)$}
\begin{center}{
\includegraphics[width=9.cm]{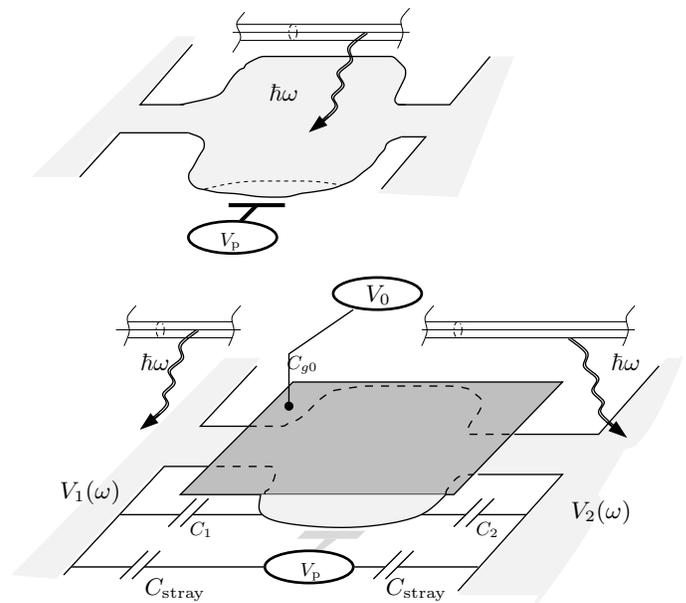}}
\end{center}}
  \caption{Top: Quantum pumping sources include oscillating voltage $V_{\rm p}(\oo)$ at
  the locally applied gate, which slightly changes the shape of the
  dot (shown dashed), or microwave antenna emitting photons with energy $\hbar\oo$ into the dot.
  Bottom: Rectification sources include external bias $V_{1,2}(\oo)$, top gate voltage
  $V_{0}(\oo)$ with capacitance $C_{g0}$, and parasitic coupling of
  $V_{\rm p}(\oo)$
  due to stray capacitances $C_{\rm stray}$. Microwave antenna can emit photons to
  the contacts and lead not only to
  photon-assisted AC transport but also to a rectified DC.}\label{fig:dot}
\end{figure}
We point here to a crucial difference between rectification and
pumping contributions to the photovoltaic effect. Rectification
results from external perturbations or the perturbations that can be
reduced to the exterior by a gauge transformation. Typical examples
are external AC-bias, or gate voltage which shifts all levels
uniformly, \cite{pedersen} or a bias induced by parasitic (stray)
capacitance which connects sources of possible internal
perturbations to macroscopic reservoirs, \cite{pump_rectif} see the
bottom panel in Fig. \ref{fig:dot}. Pumping, on the other hand, is
due to internal perturbations like those of a microwave antenna
\cite{VAA} or a locally applied gate voltage, \cite{pump} see the
top panel in Fig. \ref{fig:dot}. Internal and external sources
affect the Schr\"odinger equation and its boundary conditions,
respectively. In experiment pumping and rectification, often
considered together under the name of photovoltaic effect,
\cite{Bykov,Liu,Lin,Bartolo,BykovAB,Kvon} are hard to distinguish.

Can one clearly separate quantum pumping from rectification effects?
To distinguish them it was proposed to use magnetic field asymmetry
of DC as a signature of a true quantum pump effect. In
Refs.\,\onlinecite{pump_rectif} and \onlinecite{DiCarlo}
rectification by (non-interacting) quantum dot was due to stray
capacitances of reservoirs with pumping sources. The rectified
current was found to be symmetric with respect to $\Phi\to
-\Phi$.\cite{pump_rectif} While such field-symmetric rectification
dominated in the experiments of Switkes \etal\,\cite{Switkes} and
DiCarlo \etal\,\cite{DiCarlo} at MHz frequencies, an asymmetry
$\Phi\to -\Phi$ observed at larger GHz frequencies seemed to signify
a quantum pump effect. \cite{DiCarlo} It was noted that the Coulomb
interactions treated self-consistently do not lead to any drastic
changes in the mesoscopic distribution of a pumped
current.\cite{pump} Probably, that is why the effect of interactions
on the rectification have not been considered yet, even though the
Coulomb interaction in such dots is known to be
strong.\cite{Zumbuhl}

However, as it turned out later, Coulomb interactions are
responsible for magnetic-field asymmetry in nonlinear transport
through quantum dots. \cite{SB} Similarly this could be expected for
rectification as well. Then the magnetic field asymmetry alone can
not safely distinguish pumping from rectification. Therefore we
thoroughly examine the frequency dependence of the magnetic-field
(anti)symmetric conductances $\Ga,\Gs$. Here we neglect any quantum
pumping effects and their interference with rectification.
\cite{Vavilov05,Moskalets_AC} While the role of Coulomb interactions
and the full frequency dependence in quantum pumping are yet to be
explored, here we answer two important questions concerning a
competing mechanism, rectification: (1) In the DC limit $\oo\to 0$
for a strongly interacting quantum dot $\Ga$ and $\Gs$ are of the
same order. Is this also the case at finite frequencies? (2) How are
the experimental data affected by the bias mode for alternating
voltages?

A number of very recent experiments on nonlinear DC transport
\cite{ensslin,Bouchiat} and AC rectification
\cite{Bouchiat_preprint} have used submicron ring-shaped samples
with a relatively large aspect ratio. In this work we develop a
model of a ring which includes chaotic dynamics due to possible
roughness of its boundary and/or the presence of impurities.
Similarly to quantum dots, the two-terminal nonlinear conductance of
such a ring is field-asymmetric because field-asymmetry exists in
each arm. In particular, this leads to deviations of the phase in AB
oscillations from $0\mbox{(mod) }\pi$ which characterizes linear
conductance obeying Onsager symmetry relations. Experiments find
that the amplitude and phase of AB oscillations exhibit rather
curious properties. For example, the DC experiment of Leturcq
\etal\, \cite{ensslin} finds that during many AB oscillations with
period $hc/e$ the phase is well-defined. The experiment demonstrates
that a nearby gate can vary the phase of the AB oscillations over
the full circle. The amplitude of the second harmonic $hc/2e$ is
strongly suppressed. On the other hand, the DC experiment
\cite{Bouchiat} and AC experiment \cite{Bouchiat_preprint} of Angers
\etal\,find that the phase can be defined only for few oscillations
at low magnetic fields. For high frequencies, the phase fluctuates
strongly as function of frequency. Both in the nonlinear and the
rectified current the amplitude of the second harmonic $hc/2e$ in AB
oscillations is always comparable with the first harmonic $hc/e$.
This is in contrast with the experiments in
Ref.\,\onlinecite{ensslin}. Although we do not fully address all
these questions here, our model of a chaotic ring allows us to
consider them at least on a qualitative level.
\section{Principal results}

To introduce the reader to the problem of nonlinear transport in
Sec. \ref{sec:bias} we first qualitatively discuss the Coulomb
interaction effect in the simplest DC problem. In reality the
statistical properties of conductances $\G_{\a\b\g}$ in Eq.
(\ref{eq:IV}) are sensitive to electronic interference but to assess
the role of Coulomb interactions we can consider a specific sample.
In contrast to linear transport, it turns out that the nonlinear
current strongly depends on the way voltages at the contacts and/or
nearby conductors are varied from their equilibrium values (bias
mode). For example, we find that the experiments when only one
voltage at the contact is varied \cite{Lofgren,marlow,Bouchiat} or
when two contact voltages are shifted oppositely
\cite{Zumbuhl,ensslin} measure different nonlinear currents. Indeed,
for a current $I(\{V_i\})$, bilinear in voltages, its second
derivative should depend on the chosen direction in the space of
voltages $\{V_i\}$. Interestingly, a sample with weak interactions
is very sensitive to the choice of the bias mode, which we attribute
to the strong effect of capacitive coupling of the sample with
nearby conductors.

To make our arguments quantitative and consider the role of magnetic
flux $\Phi$ for a quantum dot which is (generally) AC-biased at
arbitrary frequency $\oo$, in Sec. \ref{sec:DC} we take electronic
interference into account. Having done that, we illustrate the
interplay between interactions and interference on several important
examples. First, we consider nonlinear transport due to a constant
applied voltage and then consider rectification of AC voltages.

For a two-terminal dot, in a generally asymmetric circuit
(capacitive couplings included), in Sec. \ref{sec:2terminal} we find
the statistics of (anti) symmetric conductances $\Ga,\Gs$ defined in
Eq. (\ref{eq:IVfield}). Both  $\Ga$ and $\Gs$ vanish on average.
Quantum fluctuations of $\Gs$ strongly depend on the interaction
strength, circuit asymmetry and bias mode. This is in accordance
with our qualitative picture. On the other hand, the antisymmetric
component $\Ga$ depends only on interactions. Our arguments agree
with recent experiments in quantum dots:\cite{Lofgren,Zumbuhl}
depending on the bias mode different features of the nonlinear
conductance tensor are probed. The fluctuations of nonlinear current
can be minimized or maximized (on average), which becomes important
for weakly interacting electrons. Curiously, for symmetric coupling
(transmission and capacitance) of contacts and dot the bias mode in
which the voltages at the contacts are changed in opposite
directions generally {\it minimizes} fluctuations of $\Gs$.
Consequently, such a mode is more advantageous for the observation
of $\Ga$ or a cleaner linear signal. Near the end of Sec.
\ref{sec:2terminal} we also demonstrate how to take into account
possible classical circuit-induced asymmetry \cite{Lofgren} due to
the finite classical resistance of the wires.

In Sec. \ref{sec:rectify} we present results elucidating the role of
interaction in rectification through two-terminal dots. Usually
there are two important time-scales: the dwell time $\dwell$ an
electron spends inside the dot and the charge relaxation time
$\RC\leq\dwell $ of the dot. For a given geometry, the dwell time
depends on the coupling of the dot with reservoirs, but the charge
relaxation time is also sensitive to the interaction strength. We
have $\RC\ll\dwell$ for strong interactions and $\RC=\dwell$ in the
weak interaction limit. Our results for fluctuations of
$\Ga(\oo),\Gs(\oo)$ are obtained for arbitrary frequency $\oo$.
Although the fluctuations of both $\Ga(\oo)$ and $\Gs(\oo)$
monotonically decrease when $\oo\to\infty$, as functions of
frequency $\oo$ they behave differently.  At nonadiabatic
frequencies $\oo\dwell\gg 1$ the nearby gate short-circuits
currents. This effect is even in magnetic field and thus affects
only $\Gs$. As a result, for a high-frequency voltage the asymptotes
of $\Ga$ and $\Gs$ are generally different and strongly depend on
the bias mode. Since the regime of parameters is quite realistic, we
expect that the predicted difference of $\Gs(\oo)$ and $\Ga(\oo)$
should be experimentally observable. In the noninteracting limit our
results qualitatively agree with those in diffusive metallic
junctions.

Our model of a ring consisting of a chaotic dot with a ballistic arm
which encloses an AB flux  is presented in Section \ref{sec:phase}.
Although it is impossible to find the full mesoscopic distribution
of the AB phase $\delta$, its shape can be discussed qualitatively.
Since $\tan\d$ is similar to the asymmetry parameter $\A=\Ga/\Gs$ in
quantum dots, its distribution can become very wide for a particular
choice of the bias mode. On average $\la\d\mbox{(mod)}\pi\ra=0$ in
our model, and we find the dependence of the fluctuations of $\d$ on
temperature, interactions, and number of channels of the contacts
and the arm. Our treatment allows a straightforward generalization
to treat AC voltages applied to the ring. The technical calculations
are presented in the Appendix.

\begin{figure}{{\psfrag{r1}{$r_1$}\psfrag{r2}{$r_2$}\psfrag{Vg0}{$V_{g0}$}\psfrag{Vg1}{$V_{g1}$}
\psfrag{Cg0}[][][0.8]{$C_{g0}$}\psfrag{Cg1}[][][0.8]{$C_{g1}$}
\psfrag{V1}{$V_1$}\psfrag{V2}{$V_2$}
\begin{center}{
\includegraphics[width=9.cm]{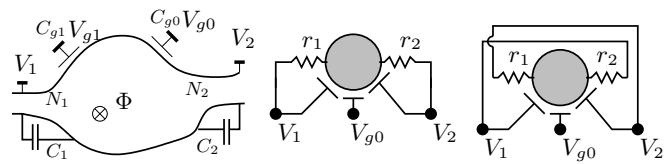}}
\end{center}}}
\caption{(Left) Rectified current is measured through a coherent
quantum dot biased by voltages with (AC) amplitude $V_{i,\oo},i=1,2$
at reservoirs connected by $N_{i}$ ballistic channels and
capacitances $C_{i}$ and by voltages $V_{gi,\oo}$ applied at
additional gates with capacitances $C_{gi}$. Transport through the
dot is sensitive to the total magnetic flux $\Phi$ through the area
of the dot. (Center and right)  Forward and reverse connection of
Ref.\,\onlinecite{Lofgren} exchange voltages at the contacts and
classical resistors $r_{1,2}$.} \label{fig:3dot}
\end{figure}
\section{Model}\label{sec:model}
The 2D quantum dot, see the left panel in Fig. \ref{fig:3dot}, is
biased with several voltages $\{V_{i}\}$ at $M$ ballistic quantum
point contacts (QPCs) with $N_{i}, i=1,...,M$ orbital channels. The
reservoirs can be capacitively coupled to the dot via capacitances
$C_i$.  An additional set of voltages $\{V_{gi}\}$ is applied to
(several) gates with capacitances $C_{gi}$. All perturbations are
assumed to be at the same frequency $\oo$, which is not necessarily
small (adiabatic).

The dot is in the universal regime, \cite{Beenakker} when the
Thouless energy $E_{\rm Th}=\hbar/\erg$ is large. The dots with area
$A=\pi L^2$ (taken circular) are either diffusive with mean free
path $l\ll L$, or ballistic, with $l\gg L$ and chaotic classical
dynamics (in the latter case the substitution $l\to \pi L/4$ should
be used). The mean level spacing (per spin direction)
$\Delta=2\pi\h^2/(m^*A)$ and the total number of ballistic channels
$N$ together define the dwell time $\dwell=h/(N\Delta)\gg \erg$. We
also require that $eV\ll N\Delta$ when we can treat the nonlinearity
only to $(eV)^2$. Scattering is spin-independent and this spin
degeneracy is accounted for by the coefficient $\nu_s$.

The noninteracting electrons are treated using the scattering matrix
approach and Random Matrix Theory (RMT) for the energy-dependent
scattering matrix $\S(\e)$. For details we refer the reader to
reviews. \cite{Beenakker,ABG} In this approach the fundamental
property of a dot is its scattering matrix $\S$ distributed over
circular ensembles of proper symmetry, see
Ref.\,\onlinecite{Beenakker} (An alternative method is the
Hamiltonian approach based on the properties of the dot's
Hamiltonian $\cal H$ taken from a Gaussian Ensemble. \cite{ABG})
Transport properties of chaotic dots in RMT for matrices $\S$ or
$\cal H$ are usually expressed in terms of an effective, magnetic
field-dependent number of channels. Predictions based on this
approach are in good agreement with experiment. For multichannel
samples with $N\gg 1$ we use the diagrammatic technique described in
Refs.\,\onlinecite{PietBeen} and \onlinecite{iop}.

However, when interactions are present, this treatment should be
modified. The approach which assumes that in a pointlike scatterer
the interactions appear in the form of a self-consistent potential
was introduced by B\"uttiker and co-authors \cite{buttiker1} on the
basis of gauge-invariance and charge conservation. This (Hartree)
approach neglects contributions leading to Coulomb blockade (Fock
terms), but is a good approximation for open systems. If the
screening in the dot inside the medium with dielectric constant $\e$
is strong, $r_s=(k_{\rm F}a_B)^{-1}=e^2/(\e\hbar v_{\rm F})\lesssim
1$, an RPA treatment of Coulomb interactions is sufficient. For
large dots, $L\gg a_B$, the details of screening potential on the
scale $\sim a_B$ are not important and we can assign an electric
potential $U(\vec r,t)$ defined by excess electrons at $\vec r,t$ at
any point $\vec r$ of the sample. If additionally the number of
ballistic channels $N$ is much smaller than the dimensionless
conductance of a closed sample, $g_{\rm dot}=\Thou/\Delta\gg N$, the
potential drops over the contacts and therefore in the interior of
the dot it can be taken uniform ("zero-mode approximation").
\cite{ABG} This potential shifts the bottom of the energy band in
the dot and thus modifies the $\S$-matrix. As a consequence,
electrons with kinetic energy $E$ have an electro-chemical potential
$\tilde E_\a=E-eV_\a$ in the contact $\a$ and $\tilde E=E-eU$ in the
dot. (We point out that we neglect the quantum pumping in the dot
and consequently the $\S$-matrix depends only on one energy.)
Recently, Brouwer, Lamacraft, and Flensberg demonstrated that this
self-consistent approach gives the leading order in an expansion in
the inverse number of channels $1/N\ll 1$. \cite{BLF} Therefore, our
analytical results present the leading order effect, valid for
$1/N\ll 1$.

In the self-consistent approach the influx of charge changes the
internal electrical potential of the dot $U(t)$, which in turn
affects the currents incoming through each conducting lead and/or
redistributes charges among the nearby conductors (gates). Such
capacitive coupling can often be estimated simply from the
geometrical configuration. For example, the capacitance of a dot
covered by a top gate at short distance $d\ll L$ is $C\sim \e L^2/d$
and a single quantum dot has $C\sim \e L$. The ratio of charging
energy $E_c\sim e^2/C$ to mean level spacing $\Delta$ characterizes
the interaction strength. It is proportional to the ratio of the
smallest geometrical scale to the effective Bohr's radius,
$E_c/\Delta\sim \mbox{min }\{d,L\}/a_B$. We refer to interactions as
strong if $E_c\gg \Delta$ and weak if $E_c\ll \Delta$.

\section{Importance of bias mode}\label{sec:bias}
\begin{figure}{{\psfrag{V1}{$x$} \psfrag{V2}{$y$}
\begin{center}{
\includegraphics[width=9.cm]{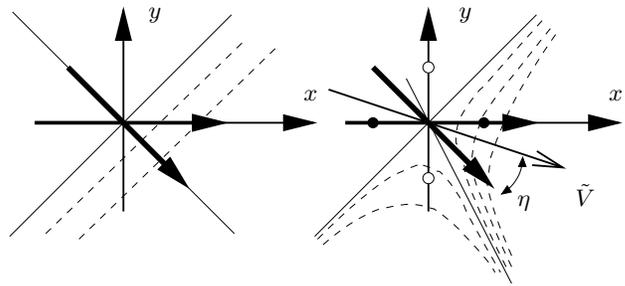}
}
\end{center}}}
  \caption{Depending on the bias mode, the experiment probes different
  transport properties. Plots present (left) linear and (right) nonlinear
  components of the current as functions of $x=V_1-V_0$ and $y=V_2-V_0$, the
  dashed curves correspond to equal currents. Thin line shows fixed $V_2$ and
  $I(\tilde V)$ is a function of source voltage $\tilde V=V_1$.
   Thick line corresponds to fixed $V_1+V_2$, such that $I(\tilde V)$ depends
   only on $\tilde V=(V_1-V_2)/\sqrt{2}$.
   Full and empty dots on the right figure correspond to the forward or reverse
   configurations shown in Fig. \ref{fig:dot}.}\label{fig:Vaxes}
\end{figure}

We suppose for simplicity that at equilibrium the voltages
$V_1=V_2=V_0$ are set. In the following we consider the situation
when the (single) gate voltage $V_0$ is held fixed at its
equilibrium value.  Experiments can be performed in different {\it
bias modes}, usually either (i) with fixed drain voltage $V_2$ or
(ii) at fixed $V_1+V_2$ (the variations of the voltages at the
contacts are equal in magnitude but opposite in sign). These
different modes correspond to straight lines in the $\{V_1 , V_2\}$
plane shown in Fig. \ref{fig:Vaxes}.

Let us consider the nonlinear current as a function $I(x,y)$, where
$x=V_1-V_0$ and $y=V_2-V_0$ are deviations of contacts voltages from
equilibrium. For generality we consider below a situation when the
linear combination $-x\sin(\eta-\pi/4)+y\cos(\eta-\pi/4)=0$ is held
fixed and the only variable is
\begin{eqnarray} \label{eq:tildeV}
\tilde
V=x\cos(\eta-\pi/4)+y\sin(\eta-\pi/4).
\end{eqnarray}
This corresponds to a rotation of the original $x , y$ axes such
that the new coordinate axis $\tilde V$ makes an angle $\eta$ with
the $y=-x$ line, as illustrated in Fig. \ref{fig:Vaxes}. The value
of $\eta$ fully characterizes the bias mode. Now the two modes
introduced above are simply (i) $\eta=\pi/4$ which implies $\tilde
V=x$; and (ii) $\eta=0$, which implies $\tilde V=(x-y)/\sqrt{2}$ and
corresponds to an asymmetric variation of the voltages.

The linear current depends only on $x-y$ (dashed lines on the left
panel in Fig. \ref{fig:Vaxes} correspond to the lines of equal
currents) and in any bias mode the measured linear current $I_{\rm
lin}$ is the same for a given $x-y$. If we consider the nonlinear
current $I$ as a function of $x,y$, it is by construction a bilinear
function of $x,y$. As in the linear case the current must vanish if
the voltages are the same and thus $I=0$ for $x-y=0$. Therefore, the
bilinear function must be of the form
\begin{eqnarray}\label{eq:simpleI}
I=I_0 \,\left[ (x+y)\cos\phi+(x-y)\sin\phi\right]\, (x -
y)\end{eqnarray} with unknown (generally fluctuating) parameters
$I_0$ and $\phi\in(-\pi/2,\pi/2]$. It is important that the
qualitative behavior of $I(x,y)$ depends on the interaction
strength: one could expect that transport depends not only on
voltages in the leads, but also on the internal nonequilibrium
potential $U$ of the sample. This potential can be found if
potentials in all reservoirs and the nearby gate are known.

In the limit of weak interactions the equilibrium point $V_0$ is
important, and if we reverse the bias voltage, $(V,0)\to (0,V)$ the
current is fully reversed, that is $\DD^2_{xx}I=-\DD^2_{yy}I$. For
the current defined in Eq. (\ref{eq:simpleI}) it is possible only
when $I\propto (x-y)(x+y)\Rightarrow \phi=0$. Another way to see
this is to use the usual expression for the total current in terms
of scattering matrices. In this formula the current depends on the
 difference between  Fermi distributions in the leads
 $\propto f(\e-ex)-f(\e-ey)$, and its expansion up to the second order yields
 $f''(\e)(x^2-y^2)$. The lines
of equal current are curved and directions $\eta=0,\pm \pi/2$
correspond to zero current directions. Thus the dependence of
current on the angle $\eta$ is strong. In addition this approach
predicts that the current through a two-terminal sample is symmetric
with respect to the magnetic flux inversion.

In contrast, for strong interactions, the value of $V_0$ is
irrelevant and the nonequilibrium electrical potential $U$ is
independent of $V_0$. In this case current depends only on the
voltage difference $x-y$ and thus $I\propto (x-y)^2\Rightarrow
|\phi|=\pi/2$. The equal-current lines are straight and the picture
is similar to the left plot in Fig. \ref{fig:Vaxes} for linear
transport. Therefore we do not expect any nontrivial dependence of
the nonlinear current on the choice of the bias mode.

It is noteworthy that qualitative considerations can predict neither
the sign, nor the magnitude of $I_0$. The only general conclusion
which we can make for a weakly or strongly interacting dot is
$I(x,y)\propto x^2-y^2$ and $I(x,y)\propto (x-y)^2$, respectively.
Experiments extract derivatives of $I$ with respect to the applied
voltages. Importantly, this derivative depends on the chosen
direction $\eta$. The nonlinear current measured in this bias mode
is
\begin{eqnarray}\label{eq:IVangle}
\label{eq:Iexample} I(\eta)&=& I_0 \tilde V^2\cos\eta\sin
(\phi+\eta).
\end{eqnarray}
The current is zero when $\eta=-\phi$ and $\eta=\pm\pi/2$, and the
bisectrix of the angle between the two zero-current directions at
$\eta=-\phi/2+(\pi/4) \mbox{sgn }\phi$ maximizes $\DD^2 I/\DD\tilde
V^2$.

Sometimes experiments extract information on nonlinearity from
measurements in different connections schematically shown in the
central and right panels in Fig. \ref{fig:3dot}: "Forward"
connection corresponds to $x=\pm V,y=0$, while "reverse" connection
for the same voltage configuration corresponds to $x=0,y=\pm V$. The
gate voltages $V_g$ are kept fixed. In Fig. \ref{fig:Vaxes} these
forward and reverse points are indicated by black and white dots,
respectively. To find the nonlinear conductance Marlow
\etal\,\cite{marlow} and L\"ofgren \etal\,\cite{Lofgren} determine
the difference of conductance at these measurement points. L\"ofgren
\etal\,\cite{Lofgren_2004,Lofgren} use the term "rigidity" for
samples for which $G_f(V)=G_r(-V)$ in the points $f^+ =(V,0)$ and
$r^- =(0,-V)$. \cite{Lofgren_2004} Equation (\ref{eq:simpleI}) gives
the nonlinear contribution $\G$ to the full conductance $G_{f,r}(\pm
V)$:
\begin{eqnarray}
\G\propto I_0\left[(x-y)\sin\phi+(x+y)\cos\phi\right]\,.
\end{eqnarray}
Thus for a sample which is called rigid this implies $I_0\cos\phi\to
0$. Since $I_0 = 0$ would mean that there is no second-order
response, we must have $\cos\phi\to 0$ which is the case for samples
with strong interaction. In other words, "rigidity" in samples which
exhibit $O(V^2)$ current is equivalent to strong Coulomb
interactions.

On the other hand, comparison of data at another pair of points
$f^+=(V,0)$ and $r^+=(0,V)$ gives $G_f(V)-G_r(V)\propto I_0\sin\phi$
and provides {\it additional} information about the two fluctuating
quantities $I_0,\phi$. Reference \onlinecite{Lofgren} expects that a
Left-Right (LR)-symmetric system has $G_f(V) = G_r(V)$. Therefore
rigid and LR-symmetric sample should necessarily have $I_0 \to 0$
and thus could not exhibit a second-order current $O(V^2)$. This
point is discussed more quantitatively in Sec. \ref{sec:2terminal}.

It is important to note that to find the linear DC current one needs
to know only $x-y=V_1-V_2$, while for the nonlinear current in
general one needs two variables $x=V_1-V_0,y=V_2-V_0$ or any
independent pair of their linear combinations. The projection of the
vector $(V_1,V_2,V_0)$ on the $V_1+V_2+V_0=$const plane uniquely
defines the nonlinear current. This projection can be parametrized
by the pair of Cartesian $(x,y)$ or axial coordinates $(\tilde
V,\eta)$. However, if in the experiment the voltages $V_{1,2}$ were
fixed, this would not be enough to define $(x,y)$ uniquely. In this
case Ref.\,\onlinecite{Lofgren} points to the importance of the
reference point $V_0$. Indeed, one could arrive at the point with a
given $(V_1,V_2)$ from any equilibrium point and the measured
current would depend on $V_0$. We prefer to characterize the
measurement by the pair $(\tilde V,\eta)$ instead of  three
variables $(V_1,V_2,V_0)$ because of the simplicity of the final
results. The weaker the interaction (or the stronger the capacitive
coupling of the sample to the nearby gate) the more important the
role of $\eta$ chosen in experiment.

We illustrate this important conclusion by quantitative results
 for nonlinear conductance $\G\propto\DD^2 I/\DD\tilde V^2$ in the following sections.
We point out that conductance with respect to the voltage difference
$V=V_1-V_2$ is often used, even when a linear combination $\tilde V$
is actually varied in experiment. Voltages $\tilde V$ and $V$ are
related, $\tilde V=V/\sqrt{2}\cos\eta$, and one can
straightforwardly find $\DD^2 I/\DD V^2$.

\section{Generation of DC in quantum dots}\label{sec:DC}

Now we quantify the qualitative arguments of Sec. \ref{sec:bias} and
consider the more general situation of a DC current generated by an
AC bias. If at first we neglect Coulomb interactions, the nonlinear
DC current $I_\a$ in response to the Fourier components
$V_{\b,\oo}=V_\b e^{i\phi_\b}$ of the AC voltages applied at the
contacts $\b=1,...,M$, can be expressed with the help of the
DC-conductance matrix $g_{\a\b}(\e)$ of the dot at the energy $\e$
\cite{pedersen}
\begin{eqnarray}\label{eq:Pedersen}
I_\a &=&\frac{\nu_s e^3}{h}\int d\e
\frac{f(\e+\hbar\oo)+f(\e-\hbar\oo)-2f(\e)}{(\hbar\oo)^2}
 \nonumber \\ &\times&
\sum_{\b=1}^M g_{\a\b}(\e)|V_{\b,\oo}|^2,\\
  \label{eq:gDC} g_{\a\b}(\e)&=&\Tr [{1\!\! 1}_\a\d_{\a\b}-\S^\dagger(\e)
  {1\!\! 1}_\a\S(\e){1\!\! 1}_\b].
\end{eqnarray}
If we now include interactions using a self-consistent potential
$U_{\oo}$ this formula is modified: \cite{pedersen} in Eq.
(\ref{eq:Pedersen}) the Fourier components of the voltages at {\it
all} contacts are shifted down by the Fourier component of the
internal potential $-U_\oo$\,
\begin{eqnarray}
U_\oo &=&\sum_\g u_\g V_{\g,\oo}, \,\,u_\g=
\frac{\sum_{\b}G_{\b\g}(\oo)-i\oo
C_\g}{\sum_{\b\g}G_{\b\g}(\oo)-i\oo
C_\Sigma}
\label{eq:uomega},\\
 G_{\b\g}(\oo)&=& \frac {\nu_s e^2}{h}\int
d\e\,\Tr\left[{1\!\! 1}_\b{1\!\! 1}_\g - {1\!\! 1}_\g{\cal
S}^\dagger(\e){1\!\! 1}_\b{\cal S}(\e+\hbar\oo)\right]\nonumber
\\ &\times&
\frac{f(\e)-f(\e+\hbar\oo)}{\hbar\oo}.\label{eq:sumG}
\end{eqnarray}
In Eq. (\ref{eq:uomega}) the index $\g$ runs not only over real
leads $1,...,M$, but also  over all gates $gi$. However, when
$\g\in\{gi\}$ the AC conductance $G_{\b\g}(\oo)$ is absent and only
capacitive coupling $i\oo C_\g$ remains in the numerator. We point
out that the matrix $ G(\oo)$ of dynamical AC conductance at
frequency $\oo$ given in Eq. (\ref{eq:sumG}) should not be confused
with the degenerate matrix $g(\e)$ of energy-dependent DC
conductances of electrons with kinetic energy $\e$ given in Eq.
(\ref{eq:gDC}).

The results of Ref.\,\onlinecite{pedersen} can be expressed in terms
of the DC conductances $g_{\a\b}$ and frequency-dependent
characteristic potentials $u_\g$,
\begin{eqnarray}\label{eq:current4omega}
I_\a &=&\frac{\nu_s e^3}{h}\int d\e
\frac{f(\e+\hbar\oo)+f(\e-\hbar\oo)-2f(\e)}{(\hbar\oo)^2} \nonumber \\
&\times &\sum_{\b\g} g_{\a\b}(\e)\mbox{Re }\, u_\g
|V_{\b,\oo}-V_{\g,\oo}|^2.
\end{eqnarray}
Here $\mbox{Re }u_\g$ stands for the real part of $u_\g$, which is
in general a complex quantity. In contrast to Eq.
(\ref{eq:Pedersen}), Eq. (\ref{eq:current4omega}) is expressed via
differences of voltages applied to all present conductors.
Therefore, the current is gauge-invariant. The charge conservation,
$\sum_\a I_\a=0$, is obvious from Eq. (\ref{eq:gDC}).

From this point on we consider Eq. (\ref{eq:current4omega}), a
specific expression of Eq. (\ref{eq:IV}), in detail for several
regimes. In Sec. \ref{sec:2terminal} we discuss the nonlinear
current due to DC applied voltages (previously considered in Ref.
\onlinecite{ChristenButtiker}) and the importance of different bias
modes in experiments in two-terminal quantum dots. In Sec.
\ref{sec:rectify} we consider the frequency dependence of $\Gs(\oo)$
and $\Ga(\oo)$.

\subsection{Nonlinearity in quantum dots}\label{sec:2terminal}

In the static limit \cite{ChristenButtiker} $\hbar\oo/T\to 0$ the
integrand in the first line of Eq. (\ref{eq:current4omega})
simplifies to $f''(\e)$ and for $\hbar\oo/N\Delta\to 0$ the
derivatives $u_\g$ are real and expressed via subtraces of the
Hermitian Wigner-Smith matrix $\S^\dagger \DD_\e\S/(2\pi i)$
\cite{WignerSmith,BP}
\begin{eqnarray}\label{eq:current4}
I_\a &=&\frac {-\nu_s e^3}{h}\sum_{\b\g}\int f'(\e)d\e
g'_{\a\b}(\e) u_{\g}(V_\b-V_\g)^2,\\
\label{eq:u} u_\g &=&\frac{ C_\g/\nu_s e^2-\int d\e f'(\e)\Tr {1\!\!
1}_\g\S^\dagger \DD_\e\S/(2\pi i) }{C_\Sigma/\nu_s e^2 -\int d\e
f'(\e)\Tr \S^\dagger \DD_\e\S/(2\pi i)}\label{eq:u0}.
\end{eqnarray}
For a two-terminal sample the nonlinear current through the first
lead is
\begin{eqnarray}\label{eq:current}
I_1 &=&\frac {-\nu_s e^3}{h}\int f'(\e)g'_{11}(\e)d\e\left[\sum_i
u_{gi}\left[(V_1-V_{gi})^2 \right.\right.\nonumber \\ && \mbox{}
\left.\left. -(V_2-V_{gi})^2\right]+(u_2-u_1)(V_1-V_2)^2\right].
\end{eqnarray}
The characteristic potentials in the last term of Eq.
(\ref{eq:current}) are sensitive to the asymmetry of the contacts.
Indeed, in a strongly interacting dot $u_{gi}=0$ and $u_2-u_1\approx
(N_2-N_1)/N$. The current magnitude grows with asymmetry due to the
last term in Eq. (\ref{eq:current}). On the other hand, the sign of
$I_1$ is random because of quantum fluctuations of $g'_{11}$ around
zero. \cite{deriv} As a consequence, if in an experiment the Fermi
level is shifted by $\d\m_{\rm F}\sim N\Delta/2\pi$ (or the shape of
the dot is changed) the sign of nonlinearity can be inverted.

Different modes of bias having been discussed in Sec.
\ref{sec:bias}, we concentrate here on the (anti)symmetric
conductances through the quantum dot at fixed gate voltages. When
the reservoir voltages are varied in the $\eta$ direction, the
nonlinear current is given by the expression
\begin{eqnarray}\label{eq:derivIV}
I &=&\frac {-2\nu_s e^3}{h}\int
f'(\e)g'_{11}(\e)d\e\left[(1-u_1-u_2)\sin \eta\right.\nonumber
\\ &&\left.+(u_2-u_1)\cos\eta\right]\cos\eta\tilde V^2,
\end{eqnarray}
and one can define exactly the unknown parameters $I_0,\phi$ which
we introduced in the qualitative argument leading to Eq.
(\ref{eq:Iexample}). Depending on $\eta$ one measures different
linear combinations of conductances. If we consider conductances
$\DD^2 I/2\DD \tilde V^2$ in units of $\nu_s e^3/h$, Eqs.
(\ref{eq:IVfield}) and (\ref{eq:derivIV}) yield
\begin{eqnarray}\label{eq:defG}
{\cal G}_{a,s}=\frac{2\pi\cos^2\eta\int d\e d{\tilde
\e}f'(\e)f'(\tilde \e)\chi_1(\e)\chi_{2,a(s)}(\tilde
\e)}{\Delta^2[C_\Sigma/(e^2\nu_s)-\int d\e f'(\e)\Tr\S^\dagger
\DD_\e\S/(2\pi i)]}
\end{eqnarray}
 expressed in terms of fluctuating functions $\chi$ and a
traceless matrix $\Lambda= (N_2/N){1\!\! 1}_1-(N_1/N){1\!\! 1}_2$:
\begin{eqnarray}\label{eq:chi1}
\chi_1(\e) &=& (\Delta/2\pi)\DD_\e \Tr\Lambda {\cal
S}^\dagger\Lambda{\cal S}, \\
\label{eq:chi2a}\chi_{2,a}(\e) &=&(i\Delta/2\pi)
\Tr\Lambda[\S^\dagger,\DD_\e\S],\\
\label{eq:chi2s}
 \chi_{2,s}(\e) &=&\Delta\left(\frac{C_0\tan\eta+C_2-C_1}{e^2\nu_s}
 +\frac{N_2-N_1}{N}\frac{\Tr \S^\dagger\DD_\e\S}{2\pi i }\right.\nonumber
\\ &&\left. +\frac{1}{2\pi i}
\Tr\Lambda\{\S^\dagger,\DD_\e\S\}\right).
\end{eqnarray}
Standard calculations using the Wigner-Smith and/or $\S$-matrix
averaging \cite{waves,PietBeen,iop} yield $\la\Ga\ra=\la\Gs\ra=0$.
This result signifies that the nonlinear current through a quantum
dot is indeed a quantum effect. As a consequence the size of the
measured nonlinearity must be evaluated from correlations of
$\Ga,\Gs$.

The functions $\chi_1(\e,\Phi)$ and $\chi_{2,a/s}(\e',\Phi')$ are
uncorrelated, and their autocorrelations \cite{PhysicaE} readily
allow one to find statistical properties of ${\cal G}_{a,s}$. Our
results can be expressed in terms of diffuson $\Diff$ or cooperon
$\Coop$ in a time representation, $\exp(-\tau/\tau_{\Diff})$ and
$\exp(-\tau/\tau_{\Coop})$. Both can be introduced using the
$\S$-matrix correlators \cite{PVB} (correlations of retarded and
advanced Green functions lead to the same expression up to a
normalization constant \cite{ABG}). We have
\begin{eqnarray}
  {\cal S}(\tau,\Phi) &=&
  \int\frac{d\e}{2\pi\hbar} \,
  {\cal S}(\e,\Phi)e^{i \e\tau/\hbar},\nonumber \\
  \langle {\cal S}_{i j}
  (\tau,\Phi)
  {\cal S}^{*}_{k l} (\tau',\Phi')\rangle
  &=& (e^{-\tau/\tau_\Diff}\delta_{ik} \delta_{jl} + e^{-\tau/\tau_\Coop}
  \delta_{il} \delta_{jk})\nonumber \\
  &\times &\frac{\Delta}{2\pi\hbar}\delta(\tau-\tau') \theta(\tau),
\label{eq:cum1t}\\
\label{eq:channels}
\tau_{\Coop,\Diff}=\frac{h}{N_{\Coop,\Diff}\Delta},\, \Brace{N_{\cal
C}}{N_{\cal D}} &=& N+\frac{(\Phi\pm\Phi')^2}{4\Phi_0^2}\frac{h v_F
l}{L^2\Delta}.
\end{eqnarray}
 We also introduce the electrochemical
capacitance $C_\m$ \cite{PietMarkus} which relates the non-quantized
mesoscopically averaged excess charge $\la Q\ra$ in the dot in
response to small shift of the voltages $\d V$ at all gates. In
addition the charge relaxation time $\RC$ of the dot is conveniently
introduced by this electrochemical capacitance and the total contact
resistance,
\begin{eqnarray}\label{eq:excess}
 C_\m= \frac{\la \d Q\ra}{\d V}=\frac{C_\Sigma}{1+C_\Sigma\Delta/(\nu_s
 e^2)},\,\,\,\RC= \frac{hC_\m}{\nu_s Ne^2}.
\end{eqnarray}
The denominator of Eq. (\ref{eq:defG}) is a self-averaging quantity,
$\la(...)^2\ra=\la (...)\ra^2=\Delta^2(C_\Sigma/C_\m)^2$. Using the
diffusons and cooperons defined in Eq. (\ref{eq:cum1t}) we find the
following correlations of $\Ga$ and $\Gs$:
\begin{eqnarray}
\label{eq:main}
&&\Brace{\la\Ga(\Phi)\Ga(\Phi')\ra}{\la\Gs(\Phi)\Gs(\Phi')\ra} =
\Brace{{\mathcal F}_\Diff-{\mathcal F}_\Coop}{{\mathcal
F}_\Diff+{\mathcal F}_\Coop+X}({\mathcal F}_\Diff+{\mathcal
F}_\Coop)\nonumber \\
&&\times \left(2\cos^2\eta\frac{2\pi}
{\Delta}\frac{C_\m}{C_\Sigma}\right)^2\frac{N_1^3
N_2^3}{N^6},\\
&& {\cal F}_{\l}=\left(\frac{\Delta T }{2\hbar^2}\right)^2\int\frac
 {\tau_\l\tau^2 e^{-\tau/\tau_\l}}{\sinh^2 \pi
T \tau/\hbar}d\tau,
\label{eq:Fraw}\\
&&\label{eq:X}X= \frac{N^2}{2N_1N_2}
\left(\frac{C_0\tan\eta+C_2-C_1}{\nu_s e^2/\Delta}
+\frac{N_2-N_1}{N}\right)^2.
\end{eqnarray}
There are two very different contributions to Eq. (\ref{eq:main}),
${\mathcal F}_{\Coop,\Diff}$ due to quantum interference and $X$
defined by the classical response of the internal potential to
external voltage. The terms denoted by ${\cal F}_{\Coop,\Diff}$ are
sensitive to temperature, magnetic field, and decoherence.
Asymptotical values of ${\cal F}$ in the low temperature, $T\ll
\hbar/\tau_{\l}$, or high temperature limits, $T\gg
\hbar/\tau_{\l}$, are ${\cal F}_\l \to 1/N_\l^2=(\tau_\l\Delta/h)^2$
and ${\cal F}_\l \to \Delta/(12 T N_\l)=\tau_\l\Delta^2/ (12 hT)$,
respectively.

The term denoted by $X$ and given by Eq. (\ref{eq:X}) contains only
quantities specifying the geometry of the sample and gates and the
bias mode. In a real experiment the coupling due to capacitances
$C_{1,2}$ is usually stronger then that of the external gates,
$C_{1,2}\gg C_0$. Symmetrization of the circuit $C_1=C_2$ can
diminish the value of $X$. If in addition $N_1=N_2$ and $\eta=0$
(used in the experiments \cite{Zumbuhl,ensslin}) we have $X\to 0$.
Thus such a symmetric setup and bias mode minimize the fluctuations
of the nonlinear current and actually would be best for an accurate
measurement of {\it linear} transport. Indeed, this regime is not
affected by the fluctuations of capacitive coupling $u_0$ of the dot
with the nearby gate and thus minimizes fluctuations of $\Gs$ around
0.

Fluctuations of $\Ga,\Gs$ are given by different expressions, see
the first line of Eq. (\ref{eq:main}), where the first term is due
to $\la\chi_{2,a}^2\ra$ or $\la\chi_{2,s}^2\ra$. Importantly,
$\la\chi_{2,s}^2\ra$ contains both quantum ${\mathcal
F}_{\l}\lesssim 1/N_\l^2$ and classical $X$ contributions. If the
classical term dominates, $X\gg 1/N^2$, the current is mostly
symmetric, $\Gs^2\gg\Ga^2$. This could be expected either for a
weakly interacting dot or a very asymmetric setup, $N_1\neq N_2$
.\cite{PhysicaE} However, if the classical term is reduced due to,
e.g., the bias mode, the fluctuations of $\Ga$ and $\Gs$ become
comparable. This experimentally important conclusion remains valid
for {\it any interaction strength}. (Particularly, it leads to a
very wide distribution of the Aharonov-Bohm phase considered in Sec.
\ref{sec:phase}.)

Experiments of Zumb\"uhl \etal\,\cite{Zumbuhl} and Leturcq
\etal\,\cite{ensslin} are performed in this regime when $\eta=0$ and
$X\to 0$. Data in Ref.\,\onlinecite{Zumbuhl}
 demonstrate that the part of the total current symmetrized with respect
 to magnetic field is by far dominated by linear conductance.
From Eq. (\ref{eq:main}) we expect mesoscopic fluctuations in linear
conductance to be $\sim N^2$ times larger then those of $\Gs\Delta$.
Thus only when the number of channels is decreased will the
nonlinear $\Gs$ become noticeable. A clear observation of $\Gs$
without linear transport contribution was performed in a DC
Aharonov-Bohm experiment by Angers \etal\,\cite{Bouchiat} in the
mode $\eta=\pm \pi/4$ (only one contact voltage was varied). This
allowed to evaluate the interaction strength from the ratio of
$\Gs/\Ga$.

Experiments of Marlow \etal\,\cite{marlow} and L\"ofgren
\etal\,\cite{Lofgren} measure the full two-terminal conductance and
extract nonlinear conductance properties related to various spatial
symmetries of the dot. Although the current through a weakly
interacting sample is field-symmetric, this is not true in general.
 Samples of Ref.\,\onlinecite{Lofgren} differ in "rigidity" and
degree of symmetry.  Rigid samples, $u_0\to 0$, with
 Left-Right(LR) and Up-Down (UD)-symmetry should have $(u_2-u_1)_s=0$ and
$(u_2-u_1)_a=0$  respectively, according to the expectations of
L\"ofgren \etal\,\cite{Lofgren} (indices $s$ and $a$ mean the
symmetric and antisymmetric part in magnetic field).

Due to quantum fluctuations, in experiment none of these
symmetry-relations can be exactly fulfilled, see Eq.
(\ref{eq:main}). According to Eq. (\ref{eq:derivIV}), the difference
in the full conductances $g=(h/\nu_s e^2)I/V$ measured between
different points probes different characteristic potentials.
Reference \onlinecite{Lofgren} defines three differences $g_{\rm
i,ii,iii}$ for three pairs of points in the forward and reverse
connection discussed after Eq. (\ref{eq:IVangle}). Using Eq.
(\ref{eq:main}) we find (i) $g_{\rm i}\equiv
g_f(V,B)-g_r(-V,B)\propto u_0$, (ii) $g_{\rm ii}\equiv
g_f(V,B)-g_f(V,-B)\propto (u_2-u_1)_a$, and (iii) $g_{\rm iii}\equiv
g_f(V,B)-g_f(-V,-B)\propto (u_0+u_2-u_1)_s$. The ensemble average of
these differences
 vanishes and their fluctuations for $C_{1,2}=0,N_1=N_2$ are given
 by
\begin{eqnarray}
\left\{\begin{array}{c}
  g_{\rm i}^2 \\
  g_{\rm ii}^2 \\
  g_{\rm iii}^2\\
\end{array}\right\}=\left\{\begin{array}{c}
  X \\
   {\mathcal F}_\Diff-{\mathcal F}_\Coop\\
  X+{\mathcal F}_\Diff+{\mathcal F}_\Coop\\
\end{array}\right\}({\mathcal F}_\Diff+{\mathcal
F}_\Coop) \left(\frac{\pi e V}
{2\Delta}\frac{C_\m}{C}\right)^2,\nonumber
\end{eqnarray}
where $X=2(C/C_\m-1)^2$ is found from Eq. (\ref{eq:X}) at $\eta=\pm
\pi/4$. In weakly interacting dots $C_\m/C\to 0$ and only
magnetic-field symmetric signals $g_{\rm i}$ and $g_{\rm iii}$
survive. In strongly interacting ("rigid") dots $C_\m/C\to 1$ and
$g_{\rm ii}$ becomes similar to $g_{\rm iii}$. We point out that
even if the rigid samples are made symmetric with respect to
Left-Right inversion, the quantum fluctuations of the sample
properties are unavoidable and $g_{\rm ii}^2\neq 0$ at $\Phi\neq 0$.
For high magnetic fields and arbitrary interactions ${\mathcal
F}_\Coop\to 0$ and experiment should observe $g_{\rm i}^2+g_{\rm
ii}^2=g_{\rm iii}^2$. Clearly fluctuations exist also for large
magnetic fields beyond the range of applicabilty of RMT.
Experimental data (see inset of Fig. 6 in Ref. \onlinecite{Lofgren})
show that $g_{\rm i}^2+g_{\rm ii}^2\sim g_{\rm iii}^2$. It is hard
to make a quantitative comparison with Refs.\,\onlinecite{Lofgren}
and \onlinecite{marlow}, since the quantum fluctuations in the
nonlinear conductance exist possibly on the background of classical
effects due to macroscopic symmetries. We expect that quantum
effects become more pronounced as contacts are narrowed.

To conclude this subsection we briefly discuss here the case of a
macroscopically asymmetric setup. If the experiment were aimed to
measure large $\Ga$ compared to $\Gs$, one would try to minimize
$\Gs$ by adjusting the setup.  Such a procedure minimizes the value
of $X$ in Eq. (\ref{eq:X}). For $C_{1,2}=0, \eta=\pi/4$ the role of
asymmetric contacts $N_1\neq N_2$ was discussed in
Ref.\,\onlinecite{PhysicaE}. Analogously, one could consider a more
general case of $C_{1,2}$ and an arbitrary bias mode $\eta$. This is
especially important if the difference $C_1\neq C_2$ can not be
neglected due to occasional loss of contact symmetry.

The results of an experiment could also be affected by the presence
of classical resistance loads $r_{1,2}$ between macroscopic
reservoirs and the dot (shown in Fig. \ref{fig:3dot}).
   Swapping of such resistances in the experiment, when connection is switched
   between "forward" and "reverse" \cite{Lofgren}
affects the voltage division between loads. If we assume the
capacitive connection of the dot and reservoirs is still the same,
the modification of the expression for $u_\g$ in Eq.
(\ref{eq:uomega}) is straightforward, $\sum_{\b}G_{\b\g}\to
\sum_{\b}G_{\b\g}/(1+r_\g \sum_{\b}G_{\b\g})$. Naturally, at large
$r_\g$, $(2 e^2 N_\g/h)r_\g \gg 1$, the main drop of the voltage
occurs over the resistor $r_\g$ and not over the QPCs. As a
consequence, if  $r_{1,2}\neq 0$, values of $u_{1,2}$ can become
unequal due to $r_1\neq r_2$ and this leads to the classical circuit
asymmetry which we do not consider here.



\subsection{Rectification in quantum
dots}\label{sec:rectify}

Here we consider the DC generated by a quantum dot subject to an AC
bias at the frequency $\oo$. In experiment at high bias frequency
$\oo\dwell\gtrsim 1$ current is usually measured at zero frequency.
In contrast, at small bias frequency $\oo\dwell\ll 1$ higher
harmonics (for instance the second harmonic $2\oo$) can be measured.
However, up to corrections small due to $\oo\dwell\ll 1$, the second
harmonic is just equal to the rectified current, $I_{2\oo}\approx
I_0$. Therefore, to leading order, our results for the rectified
current describe both experiments.

Generally, there are several important time-scales characteristic
for time-dependent problems in chaotic quantum dots. To see how they
appear let us first consider frequency-dependent linear transport of
noninteracting electrons. Its statistics usually depend only on the
flux-dependent time scales $\tau_{\Coop,\Diff}$, see Eq.
(\ref{eq:channels}). If we consider an analog of UCF $\la
G^2(\Phi)\ra$ for the frequency-dependent conductances introduced in
Eq. (\ref{eq:sumG}), we find
\begin{eqnarray}\label{eq:linear}
\la G(\oo,\Phi)G(\oo',\Phi')\ra=\left(\frac{\nu_s
e^2}{h}\frac{N_1N_2}{N}\right)^2\sum_{\l=\Coop,\Diff}&& \nonumber \\
\left(\frac{\Delta T}{2\hbar^2}\right)^2\int \frac{\tau_\l
d\tau}{\oo\oo'}\frac{
e^{-\tau/\tau_\l}(e^{i\oo\tau}-1)(1-e^{i\oo'\tau})}
{[1-i(\oo+\oo')\tau_\l/2]\sinh^2 \pi T\tau/\hbar}.&&
\end{eqnarray}
The presence of $i\oo\tau_\l$ in the diffuson and cooperon
contributions in the second line of Eq. (\ref{eq:linear}) is due to
the energy dependence of the scattering matrix $\S(\e)$, which
usually brings up imaginary corrections to the matrix-element
correlators.

In a DC-problem $\oo\to 0$ it is usually useful to introduce a
dimensionless number of channels $N_{\Coop,\Diff}$ modified by the
magnetic field, see Eq. (\ref{eq:channels}). In this limit at $T\to
0$ the integration in Eq. (\ref{eq:linear}) becomes straightforward
and summation is then performed over $N_\l^{-2}$. For equal magnetic
fields, $\Phi=\Phi'$, we have $N_\Diff=N$, but $N_\Coop$ is strongly
modified by large fields, $N_\Coop\to \infty$, which suppresses the
weak localization correction and diminishes UCF. However, for an AC
problem (especially for  $\oo\dwell\gtrsim 1$) it is more convenient
to express results in terms of dimensionless quantities
$\oo\tau_{\Coop},\oo\tau_{\Diff}$. For example, from Eq.
(\ref{eq:linear}) the statistics of conductance can be easily
evaluated: $\la |G(\oo,\Phi)|^2\ra/\la G(0,\Phi)^2\ra\sim
1/(\oo\dwell)$ and the real and imaginary parts of conductance are
similar and uncorrelated at high frequency $\oo\dwell\gg 1$.

Inclusion of interactions introduces an (additional) dependence on
$\RC$, the charge-relaxation time defined in Eq. (\ref{eq:excess}).
To leading order in $1/N\ll 1$ the effect of interactions is often
to substitute $\dwell\to \RC$ in the noninteracting results, e.g.,
for the linear conductance \cite{PietMarkus} or shot noise.
\cite{BP,Hekking} Interestingly, the subleading corrections depend
on both $\RC$ and $\dwell$, e.g., in the weak localization
correction in the absence of magnetic field.\cite{PietMarkus} When
the magnetic field is increased to values which finally break
time-reversal symmetry, the appearance of different time scales
$\tau_{\Coop,\Diff}$ is expected, see e.g., Eq. (\ref{eq:linear}).
Therefore at intermediate magnetic fields, when $\tau_\Diff\neq
\tau_\Coop$, and the interactions taken into account, $\RC\neq
\dwell$, the solution of an AC problem is expected to show a
complicated dependence on all these time scales.

Indeed, if we consider the rectified current such an interplay
between $\tau_{\Coop,\Diff}$ and $\RC$ does appear. We find
$\la\Ga\ra=\la\Gs\ra=0$ and present below results for correlations
of
 $\Ga$ and $\Gs$:
\begin{eqnarray}
\Brace{\la\Ga(\Phi)\Ga(\Phi')\ra}{\la\Gs(\Phi)\Gs(\Phi')\ra}=
\Brace{{\cal F}_{U,\Diff}(\oo)-{\cal
F}_{U,\Coop}(\oo)}{{\cal F}_{U,\Diff}(\oo)+{\cal F}_{U,\Coop}(\oo)+X(\oo)}\nonumber \\
\times [{\cal F}_{G,\Diff}(\oo)+{\cal F}_{G,\Coop}(\oo)]\left(
\frac{4\pi\cos^2\eta}
{\Delta}\frac{C_\m}{C_\Sigma}\right)^2\frac{N_1^3 N_2^3}{N^6}
\label{eq:varGs}.
\end{eqnarray}
Here the functions ${\cal F}_{U}(\oo),{\cal F}_{G}(\oo)$ are
finite-frequency generalizations of Eq. (\ref{eq:Fraw})
\begin{eqnarray}
\label{eq:FU}{\cal F}_{U,\l}(\oo)&=&\left(\frac{\Delta T}{\hbar^2
\oo}\right)^2\int \tau_\l d\tau\frac{e^{- \tau/\tau_{\l}}\sin^2
\oo\tau/2 }{2\sinh^2 \pi T\tau/\hbar}\frac{1}{1+\oo^2\RC^2}
\nonumber
\\&\times&\left(1+ \mbox{Re }\frac{1+i\oo\RC}{1-i\oo\RC}
\frac{e^{i\oo\tau}}{1-i\oo\tau_\l}\right)
,\\
{\cal F}_{G,\l}(\oo)&=& \left(\frac{\Delta T}{\hbar^2
\oo^2}\right)^2\int \frac{2d\tau}{\tau_\l}\frac{e^{-
\tau/\tau_{\l}}\sin^4 \oo\tau/2}{\sinh^2 \pi T\tau/\hbar}
\label{eq:FG}.
\end{eqnarray}
 The subscripts $U(G)$ of ${\mathcal F}_{U(G)}(\oo)$ illustrate
the origin of these functions: they result from averaging of
different scattering properties over the energy band defined by
$\mbox{max }\{\hbar\oo, T,\hbar/\tau_{\Coop(\Diff)}\}$. The function
${\mathcal F}_{U}(\oo)$ is a characteristic of the internal
potential $U_\oo$, see Eq. (\ref{eq:uomega}). The function
${\mathcal F}_{G}(\oo)$ results from the energy averaging of the DC
conductance $g(\e)$. Such averaging appears because both $G(\oo)$
defined in Eq. (\ref{eq:sumG}) and $g(\e)$ in Eq.
(\ref{eq:current4omega}) are coupled to the Fermi distribution.

 The function $X(\oo)$ is
\begin{eqnarray}\label{eq:Xgeneral}
  X(\oo) &=&\frac{N^2}{2N_1N_2}
\left(\frac{(C_0\tan\eta+C_2-C_1)(1+\oo^2\dwell\RC)}{(1+\oo^2\RC^2)\nu_s
e^2/\Delta}\right.\nonumber \\ && \mbox{} \left.
+\frac{N_2-N_1}{N(1+\oo^2\RC^2)}\right)^2,
\end{eqnarray}
 and in the static limit $\oo\to 0$ it is given by Eq. (\ref{eq:X}).
 We point out that when the interactions are negligible, $E_c\sim
e^2/C\ll\Delta$, the role of the bias mode is significant. A quantum
dot with (fully broken) time-reversal symmetry can be labeled by
Dyson symmetry parameter ($\b=2$) $\b=1$. When the setup is ideal,
$C_{1,2}=0$, and $\eta\neq 0$, the fluctuations of $\Ga,\Gs$ at
large frequencies $\oo\dwell\gg 1$ are
\begin{eqnarray}\label{eq:FK}
\d\Gs &=& \la\Gs^2\ra^{1/2}=\frac{N_1 N_2}{N^2} \left(\frac
2\beta\frac{\pi}{\oo\dwell}\right)^{1/2}\frac{2\sin
2\eta}{\hbar\oo},\\
\d\Ga &=&\left(\frac{N_1 N_2}{N^2}\right)^{3/2}\frac{\nu_s
e^2}{2C}\frac{\cos^2\eta}{\hbar^2\oo^3\dwell},\,\,\b=2.\label{eq:FKasym}
\end{eqnarray}
In chaotic quantum dots the role of the Thouless energy $\Thou$ of
the open systems is often played by the escape rate $\hbar/\dwell$.
If we take this into account, our result (\ref{eq:FK}) qualitatively
agrees with that of Falko and Khmelnitskii \cite{FK} obtained for
open diffusive metallic junctions. However, when $\eta\to 0$, the
fluctuations of $\Gs$ are much smaller and for
$|N_1-N_2|\ll\oo\dwell$ they become comparable with those of the
antisymmetric conductance (\ref{eq:FKasym}).

However, very often experiments are performed in samples, where the
interaction is not weak. Since $\Ga$ and $\Gs$ are comparable for
strong Coulomb interactions in the static limit $\oo\to
0$,\cite{PhysicaE} we concentrate here on this experimentally
relevant regime of $\Delta/E_c\sim \RC/\dwell\ll 1$ and take an
ideal symmetric setup, $N_1=N_2$ and $C_{1,2}=0$. Then we have
\begin{eqnarray}\label{eq:Xomega}
X(\oo)\approx
2\tan^2\eta\left(\frac{\dwell^{-1}+\oo^2\RC}{\RC^{-1}+\oo^2\RC}\right)^2.
\end{eqnarray}
Below we consider in detail the case $\eta\neq 0$ and how this bias
mode affects the behavior of $\Gs^2(\oo)$. Several frequency regimes
can be separated: adiabatic $\oo\tau_\l\ll 1$, intermediate, where
$1/\tau_\l\ll \oo\ll 1/\RC$, and high frequencies $\oo\RC\gtrsim 1$.
The asymptotes of the functions defined in Eqs. (\ref{eq:FU}),
(\ref{eq:FG}), and (\ref{eq:Xomega}) in these regimes are presented
in Table \ref{tab:table} for reference.

For adiabatic frequencies $\oo\tau_\l\ll 1$ the integrands in Eqs.
(\ref{eq:FU}) and (\ref{eq:FG}) do not oscillate on the short time
scale $\tau_\l$. At such small frequencies ${\cal F}_U(\oo)={\cal
F}_G(\oo)$ are equal to ${\cal F}$ of Eq. (\ref{eq:Fraw}) and
$X(\oo)\propto (\RC/\dwell)^2\ll 1$ can be neglected.  This is
essentially the zero frequency regime considered before for
nonlinear DC transport.

As the frequency grows, an intermediate regime is reached when max
$\{T,\hbar/\tau_\l\}\ll \hbar\oo\ll \hbar/\RC$ and  ${\cal
F}_{U}(\oo), {\cal F}_{G}(\oo)$ start to differ.  The scattering
properties at large energy difference $\hbar\oo\gg\hbar/\tau_\l$ are
uncorrelated and the response of the dot is randomized. Therefore
both the conductance averaged over a large energy window $\hbar \oo$
and the response of the internal potential $U_\oo$ to the AC voltage
at $\oo\dwell \gg 1$ are strongly suppressed, see Table
\ref{tab:table}. As a result, if $X(\oo)$ is still negligible, both
$\Ga^2$ and $\Gs^2$ decrease with growing frequency as $1/\oo^4$.

\begin{table}
\caption{Asymptotes of ${\mathcal F}_{U,\l}(\oo),{\mathcal
F}_{G,\l}(\oo),X(\oo)$ at $T\to 0$ \label{tab:table}}
\begin{tabular}{|c|c|c|c|}
  \hline
   & Adiabatic & Intermediate & High \\
 $\mbox{Function}$ & $\oo\ll\tau_\l^{-1}$ & $\tau_\l^{-1}\ll\oo\ll \RC^{-1}$ & $\oo\RC\gg 1$ \\
  \hline
  ${\mathcal F}_U(\oo)\times N_\l^2$ & 1 & $\pi/(4\oo\tau_\l)$ &
   $\pi/(4\oo^3\tau_\l\RC^2)$ \\ \hline
  ${\mathcal F}_G(\oo)\times N_\l^2$ & 1 & \multicolumn{2}{|c|}{
  $(\pi/\oo\tau_\l)^3 $ }\\ \hline
  $X(\oo)$ & \multicolumn{2}{|c|}{$2\tan^2\eta(\RC/\dwell)^2$} & $2\tan^2\eta$ \\
  \hline
\end{tabular}
\end{table}
One could expect that interactions qualitatively change the behavior
of $\Ga,\Gs$ when the frequencies become comparable to $1/\RC\sim
NE_c/\hbar$, the scale defined by the interaction strength. At such
frequencies the response of a dot to the potentials at the contacts
is not resistive as occurs at low frequencies, but mostly
capacitive. If the frequency is high, $\oo\RC\gtrsim 1$, we have
$\mbox{Re }u_{1,2}\to 0$ and the function ${\mathcal F}_{U}$ in Eq.
(\ref{eq:FU}) is suppressed $\sim 1/(\oo\RC)^{2}$. As a consequence,
$\Ga^2$ is suppressed stronger then $1/\oo^4$ and goes as $1/\oo^6$
at $\oo\RC\gtrsim 1$. However, a more important signature of this
capacitive coupling is the growth of $X(\oo)$ in Eq.
(\ref{eq:Xomega}), which affects $\Gs^2$.

To see the role of this growth we consider now sufficiently large
fields $\Phi=\Phi'$ when only the diffuson contribution survives.
The growth of $X(\oo)$ in Eq. (\ref{eq:Xomega}) reflects enhanced
sensitivity of the internal
 potential $U_\oo$ to the gate voltage,
$X(\oo)\propto (\tan\eta \mbox{Re }u_{0})^2$. At high frequencies
the impedance of the capacitor $C$ becomes negligible and therefore
the internal potential follows the gate voltage and not the
reservoir voltages, $u_0\to 1,u_{1,2}\to 0$. Enhanced from its small
static value $\RC/\dwell$ to 1 at large frequencies, such coupling
affects the fluctuations of $\Gs(\oo)$ if $\eta\neq 0$. The
situation is somewhat similar to the weak interaction limit, when
the coupling with nearby gates was strong, $u_0\to 1,u_{1,2}\ll 1$,
and lead to $\Gs\gg\Ga$.

\begin{figure}{\psfrag{Omega}{$\oo\dwell$}
\psfrag{Ga2}{$\Ga^2$}\psfrag{Gs2}{$\Gs^2$}\psfrag{oo3}{$1/\oo^3$}\psfrag{oo6}{$1/\oo^6$}
  \includegraphics[width=9cm]{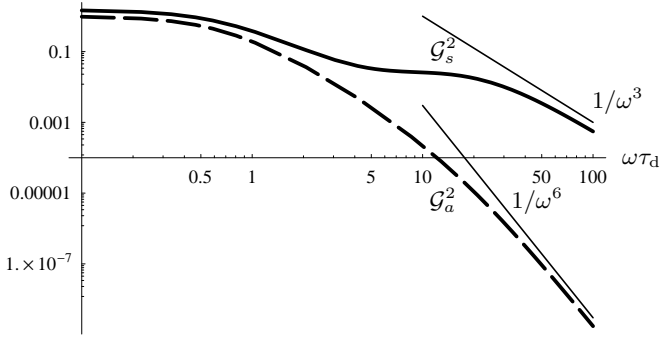}}\\
  \caption{Zero-temperature large-field fluctuations of $\Ga(\oo)$ (dashed)
  and $\Gs(\oo)$ (solid curve) in units of
  $(\pi/4\Delta N^2)^2$ for the bias mode $\eta=\pi/4$. Data are presented in the log-log scale
  at $N_{1,2}=5$ and $\RC/\dwell=0.05$.
  The asymptotes $\Ga^2\propto \oo^{-6}$ and $\Gs^2\propto
  \oo^{-3}$ are different due to $\eta\neq 0$,
  see Eqs. (\ref{eq:Gainter}) and (\ref{eq:Gsinter}). }\label{fig:GsGaMath}
\end{figure}
The fluctuations of $\Ga(\oo),\Gs(\oo)$ for $\oo\dwell\gg 1$ can be
evaluated:
\begin{eqnarray}\label{eq:Gainter}
&&\Ga^2(\oo)\sim \frac{\Delta^2}{(\hbar\oo)^4(1+\oo^2\RC^2)},\\
&&\Gs^2(\oo)\sim\Ga^2(\oo)+
\frac{(\RC\tan\eta[1+\oo^2\dwell\RC])^2}{\hbar^2\oo\dwell(1+\oo^2\RC^2)^3}.
\label{eq:Gsinter}
\end{eqnarray}
Fluctuations of $\Ga^2(\oo)$ and $\Gs^2(\oo)$ demonstrate
qualitatively different behavior, which we illustrate in Fig.
\ref{fig:GsGaMath}. Indeed, at sufficiently high frequencies, the
dependence of $X(\oo)$ on $\omega$ makes the last term in Eq.
(\ref{eq:Gsinter})  dominant. At $\oo\RC\gg 1$ the asymptotes of
  $\Ga^2\propto 1/\oo^6$ and $\Gs^2\propto 1/\oo^3$ become different due to
  the presence of the second term in Eq. (\ref{eq:Gsinter}). These results show that for
nonadiabatic frequencies of the external bias the DC current
strongly depends on the bias mode $\eta$. We predict that the
magnetic field asymmetry of the rectified current, noticeable at
small frequencies, might become suppressed for large frequencies,
when the symmetrized component dominates due to the presence of
capacitive coupling. For convenience, the low-temperature estimates
for $\la\Ga^2\ra$ and $\la\Gs^2\ra$ for $\eta\neq 0$, $\Phi\gg
\Phi_c$  are collected in Table \ref{tab:GaGs}.

\begin{table}
\caption{Estimates for $\la\Ga^2\ra$ and $\la\Gs^2\ra$,
($z=\oo\dwell$) \label{tab:GaGs}}
\begin{tabular}{|c|c|c|c|}
  \hline
   & Adiabatic & Intermediate & High \\
 $\mbox{Function}$ & $z\ll 1$ & $1\ll z\ll \dwell/\RC$ & $z\gg \dwell/\RC$ \\
  \hline
  $\frac{\hbar^4}{\Delta^2\dwell^4}\Ga^2$ & $1$ & $z^{-4}$ &
   $ \frac{\dwell^{2}}{\RC^{2}}z^{-6}$ \\ \hline
  $\frac{\hbar^2}{\RC^2}(\Gs^2-\Ga^2)$ & $1$
  & $(1+\frac\RC\dwell z^2)^2 z^{-1} $& $\frac{\dwell^4}{\RC^4}z^{-3}$ \\
  \hline
\end{tabular}
\end{table}
It is noteworthy that a recent experiment in AB rings
\cite{Bouchiat_preprint} finds that $\G(\oo,\Phi=0)$  grows with
frequency until $\oo\sim 2\Thou$ and then decreases $\sim
1/\oo^{3/2},\oo\to\infty$. While we predict a monotonic decrease of
$\la\Gs^2(\oo)\ra$, this growth could be the result of quantum
pumping or an interference of the pumping and rectification (both
effects were neglected here).


\section{Phase of Aharonov-Bohm oscillations}\label{sec:phase}

 In this section we consider nonlinear transport through a chaotic
Aharonov-Bohm (AB) ring. The nonlinear conductance $\G$ exhibits
periodic AB oscillations and non-periodic fluctuations, similarly to
the linear conductance $G$. However, since Coulomb interactions
produce asymmetry of $\G$ with respect to magnetic field inversion,
the phase of these oscillations is not pinned to $0\mbox{
(mod)}\pi$. As a quantum effect this AB phase is characterized by a
mesoscopic distribution. The width of this distribution represents a
typical fluctuation. We first discuss what kind of distribution
could be expected in a chaotic AB ring and then calculate the
fluctuation of the AB phase.

Let us assume that $\G$ as a function of magnetic flux $\Phi$ can be
expanded into the series of well-defined Fourier harmonics similarly
to the linear conductance $G$:
\begin{eqnarray}\label{eq:expansion}
\Brace{ G(\Phi)}{\G(\Phi)}&=&\sum_{n=0}^\infty
\Brace{G_n}{\G_n}\cos\left( \frac{2\pi
n\Phi}{\Phi_0}+\Brace{0}{\delta_n}\right).
\end{eqnarray}
The phase $\d$ of the main (first) harmonic $\Phi_0=hc/e$ is
obtained from the ratio of the (anti) symmetrized conductances
defined in Eq. (\ref{eq:IVfield})
\begin{eqnarray}\label{eq:tan}
\tan\delta=\frac {\int d\Phi \exp(2\pi i \Phi/\Phi_0)\Ga(\Phi)}{\int
d\Phi \exp(2\pi i \Phi/\Phi_0)\Gs(\Phi)}.
\end{eqnarray}
We can not find the full mesoscopic distribution of the phase
$P(\d)$. We can gain some insight in the behavior of this phase by
investigating a similar quantity, namely, the asymmetry parameter
$\A=\Ga/\Gs$ considered previously for chaotic dots. \cite{PhysicaE}
Based on Eq. (\ref{eq:tan}) we argue that the statistical properties
of $\arctan \A$ and the AB phase $\d$ should be similar.

In quantum dots the parameter $\A$ is given by the ratio
$\A=\Ga/\Gs=\chi_{2a}/\chi_{2s}$, see Eqs. (\ref{eq:defG}),
(\ref{eq:chi2a}), and (\ref{eq:chi2s}). The functions $\chi_{2a,2s}$
at $T\neq 0$ are convolved separately with $f'(\e)$, and at $T=0$
(which we consider below) they are evaluated at the Fermi energy.
The properties of $\chi_{2a,2s}$ and the dependence of $\chi_{2s}$
on the bias mode were described after Eq. (\ref{eq:X}). The function
$\chi_{2s}$ can have a nonzero (classical) average
$\la\chi_{2s}\ra\sim X^{1/2}$ defined by the interaction strength,
geometry of the setup, and the bias mode $\eta$. Since
$\la\chi_{2,a}\ra=0$ and the fluctuations of $\chi_{2a,2s}$ are
small as $1/N^2$, the mesoscopic distribution of $\arctan\A$ is
narrow and concentrated close to 0. However, $\la\chi_{2s}\ra=0$
 is possible if $X\to 0$, e.g., for symmetric contacts and
 the bias mode $\eta=0$. In this case, the
distribution of $\arctan \A$ becomes wide regardless of the
interaction strength.

\begin{figure}{\psfrag{x}[t]{$\phi$}\psfrag{P(x)}[l]{$P(\phi)$}
\psfrag{16-8}[l][][0.8]{$N=16,N_L=8$}
\psfrag{16-4}[l][][0.8]{$N=16,N_L=4$} \psfrag{N=2}[l][][.7]{$N=24$}
\psfrag{N=24}[l][][.7]{$N=2$}
  \includegraphics[width=8cm]{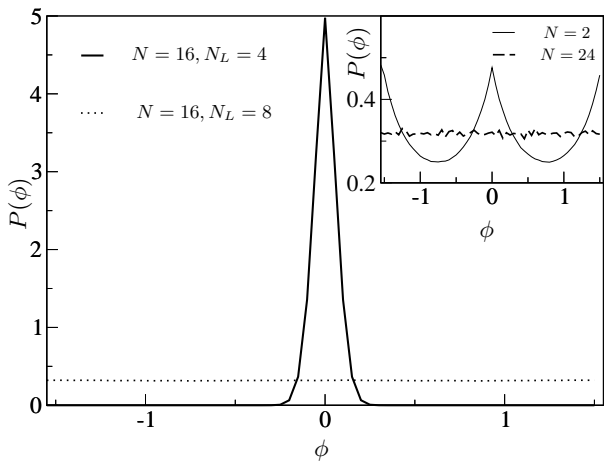}}\\
  \caption{Mesoscopic distribution $P(\phi)$ of $\phi=\arctan \Ga/\Gs$. (Main plot)
  If the contacts are asymmetric (bold curve, $N=16,N_L=4$) the distribution
  is narrow, while for symmetric contacts (dashed, $N=16,N_L=8$) it is almost
  uniform. As shown in the inset, for symmetric contacts at large $N$ the
  distribution becomes uniform, compare bold curve for $N=2$ and dashed for
$N=24$.}\label{fig:delta}
\end{figure}
The role of the classical contribution on the shape of
$P(\arctan\A)$ is demonstrated in the main plot in Fig.
\ref{fig:delta} for $\eta=0$, where the distributions for
asymmetric, $N_L=4,N=16$, and symmetric contacts, $N_L=8,N=16$, are
presented. While the distribution is almost uniform, when the
classical contribution $X$ is absent, it is highly peaked near zero
when $X$ dominates. If $X$ is absent, the correlations between $\Ga$
and $\Gs$ are significant at small $N$. This leads to a nonuniform
distribution of $P(\arctan\A)$, which is peaked at $0$ and $\pi/2$
when $N=2$, see the inset in Fig. \ref{fig:delta}. When $N$ grows,
the correlations between $\Ga$ and $\Gs$ vanish and therefore the
distribution becomes uniform. Such a distribution could be easily
obtained if we make the natural assumption that $\Ga,\Gs$ are
independent and distributed by the Gaussian law with the same width.

These numerics were performed for $\eta=0$, when the mesoscopic
distribution of $\A=\Ga/\Gs$ becomes insensitive to the interaction
strength. The role of interactions appears only if $\eta\neq 0$,
when the classical contribution $X$ becomes dominant. Similarly, we
expect that the distribution of the phase of AB oscillations is also
strongly affected by the bias mode. If the bias mode is chosen such
that the classical contribution $X$ vanishes, the phase $\d$
strongly fluctuates {\it even for weak interactions}. It would be
very interesting to check this surprising conclusion experimentally.

Let us now consider the fluctuations of the AB phase.
 Since the scattering theory turned out to
be very useful for the discussion of the nonlinear/ rectified
current through a chaotic quantum dot, we extend this theory to
rings. We make two key assumptions (discussed in the Appendix in
more detail) that the magnetic flux through the annulus of the ring
is smaller then the flux quantum $\Phi_0$ and that the mean free
path $l$, the radius $R$, the width of the ring $W$ and the contacts
$W_c$ satisfy the condition $\pi^2 l W\gg 2R W_c$. In this case the
RMT can be applied to such chaotic rings as well. Unlike the
experiments on large open rings with high aspect ratio $R/W\gg
1$,\cite{WW,Liu,Lin,Bartolo,BykovAB} the recent experiments
\cite{ensslin,Bouchiat,Bouchiat_preprint} are performed in rings of
submicron size, which are effectively zero-dimensional. The
treatment of such rings is similar to chaotic quantum dots, and the
fluctuations of $\Gs,\Ga$ can be expressed in terms of the diffuson
$\Diff$ and the cooperon $\Coop$, see Eq. (\ref{eq:main}). The only
problem is to find the expression for the effective number of
channels as a function of magnetic field, similar to Eq.
(\ref{eq:channels}).

\begin{figure}{{\psfrag{S}{$\,$}\psfrag{U}{$\,$}\psfrag{3}{$N_3$}
\psfrag{4}{$N_4$}\psfrag{1}{$N_1$}\psfrag{2}{$N_2$}
  \includegraphics[width=6cm]{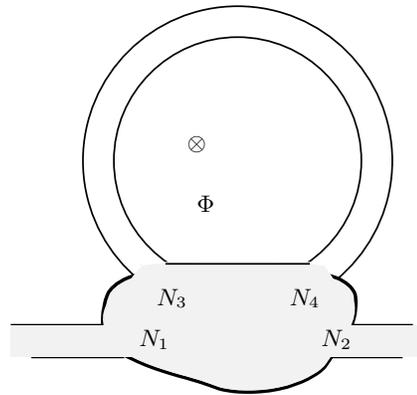}}\\
  \caption{Model of a chaotic Aharonov-Bohm ring with
  $N=N_1+N_2$ channels. The model consists of a quantum dot
 with $M$ channels combined with a ballistic arm
  with $N_3=N_4=(M-N)/2$ channels.}\label{fig:model}}
\end{figure}

The model we propose for a chaotic AB ring combines chaos and a ring
geometry: a chaotic dot is attached to a long ballistic arm which
serves to include an AB flux large compared to the fraction of the
flux through the sample. This model is shown in Fig.
\ref{fig:model}, where the ring with $N=N_1+N_2$ ballistic channels
in the contacts 1,2 is modeled by a dot with $M>N$ channels and a
ballistic arm with $N_3=N_4=(M-N)/2$ channels in contacts 3,4. The
parameter $\rho=1-N/M$, the ratio of $N_3+N_4$ to the total number
of channels $M$, can vary between 0 when the arm is much narrower
then the contacts and 1 in the opposite limit. The electronic phase
is randomized in the quantum dot, but when electrons propagate along
the arm their phase is determined by the geometry and applied
magnetic field. This model is a reasonable approximation for the
real experiment, it takes into account the long time spent by
electron inside the ring and the randomness of its motion. The
discussion of the model and the details of calculation of
$\Coop,\Diff$ are presented in the Appendix.

In experiment the Fourier transform is often taken over the total
flux (or applied magnetic field) and the flux through the hole
$\Phi_{h}$ cannot be separated from the flux through the dot
$\Phi_{d}$. Then the dependence of the diffuson and cooperon on
magnetic field is non-periodic, which is indeed observed in the form
of nonperiodic fluctuations in the (non-)linear conductance and
phase slips of AB oscillations. A possible weakness of this model is
in its spatial separation of chaotic scattering and the main part of
magnetic field, but in the limit when the arm is much wider then the
contacts $1$ and $2$ such a separation is not important and the
averaged properties of AB oscillation phase become independent of
the arm's width.

If the flux $\Phi_d$ through the dot is much smaller then the flux
$\Phi_h$ through the hole, the nonperiodic fluctuations and the
periodic AB oscillations are well-separated, which is usually the
case in experiment.\cite{ensslin,Bouchiat} In view of this
separation we can neglect the flux through the chaotic dot,
$\Phi_d\ll\Phi_h$, to find the statistics of the AB phase. We assume
that the averaging is taken over a magnetic field range containing
many AB oscillations but still small compared to the characteristic
field of the nonperiodic fluctuations. In such a simplified model of
a chaotic AB ring $N_{\Coop,\Diff}$ are given by Eq.
(\ref{eq:DCfinal}) with $\Phi_d=0,\Phi_h=\Phi$ and the parameter
$\rho=(M-N)/M$. The effective number of channels is
\begin{eqnarray}\label{eq:DCsimple}
\Brace{N_{\cal C}}{N_{\cal D}} &=& M\left(1-\rho\cos
\frac{2\pi(\Phi\pm\Phi')}{\Phi_0}\right).
\end{eqnarray}
 Using this expression for the $N_{\Coop,\Diff}$ we can evaluate
the quantum fluctuations of linear conductance in AB rings. At low
temperature a typical fluctuation of $G$ at $\Phi=0$ is $\d
G=\sqrt{2}(N_1 N_2/N^2)(\nu_s e^2/h)$ and the amplitude of AB
oscillations is $\d G_{\rm AB}^2\sim\d G^2
\rho(1-\rho)^{1/2}/(1+\rho)^{3/2}$, which reaches maximum when the
widths of the arm and the contacts are equal, $\rho=1/2$.

The first two moments of $\tan\d$ can be found analytically. It is
zero on average, $\la\tan\d\ra=0$, since the numerator and
denominator in Eq. (\ref{eq:tan}) are independent random quantities.
 Equation (\ref{eq:DCsimple})
show that $\Diff,\Coop$ are the same functions of $\Phi\pm\Phi'$, so
that all necessary
 ingredients can be expressed in terms of the functions ${\mathcal
 F}_{U,\Diff}$ and ${\mathcal F}_{G,\Diff}$ of $\Phi-\Phi'$ defined
 in Eqs. (\ref{eq:FU}) and (\ref{eq:FG}). We omit now the index $\Diff$ for
 brevity and denote the average over magnetic field by
 $\overline{ (...)}=\int_0^{\Phi_0}(...) d\Phi/\Phi_0 $,
to find
\begin{eqnarray}\label{eq:tandelta}
&&\la\tan ^2\delta\ra = \left[\overline{{\cal F}_{U }(\Phi){\cal
F}_{G }(\Phi)\cos\Phi} +\overline{ {\cal F}_{G
}}\cdot\overline{{\cal F}_{U }(\Phi)\cos\Phi}\right.\nonumber \\ &&
\mbox{} \left.-\overline{ {\cal F}_{U }}\cdot\overline{{\cal F}_{G
}(\Phi)\cos\Phi}\right]/ \left[\overline{({\cal
 F}_{U }(\Phi)+X(\oo)){\cal F}_{G }(\Phi)\cos\Phi}
 \right.\nonumber \\ && \mbox{} \left.+
\overline{\frac{{\cal F}_{U }(\Phi)}{2}}\overline{{\cal
 F}_{G }(\Phi)\cos\Phi}+
 \overline{\frac{{\cal F}_{G }(\Phi)}{2}}\overline{{\cal
 F}_{U }(\Phi)\cos\Phi}\right],
\end{eqnarray}
where the function $X(\oo)$ is defined by the setup geometry in Eq.
(\ref{eq:Xomega}). Again, in the static limit $\oo\to 0$ we have
${\cal F}_{U}={\cal F}_{G}={\cal F}$  and $X$ defined by Eqs.
(\ref{eq:Fraw}) and (\ref{eq:X}). In this case Eq.
(\ref{eq:tandelta}) can be rewritten as
\begin{eqnarray}\label{eq:tan^2result}
\frac{1}{\la\tan ^2\delta\ra} &=&1+\frac{[\overline{{\cal
 F}(\Phi)}+X]\overline{\cos\Phi{\cal
 F}(\Phi)}}{\overline{\cos\Phi{\cal
 F}^2(\Phi)}}.
\end{eqnarray}
In
the limits of high, $T\gg N\Delta/2\pi$, and low temperature, $T\ll
N\Delta/2\pi$ the asymptotical values of $\la\tan^2\d\ra$ are
\begin{eqnarray}\label{eq:tan^2asymp}
\frac{1}{\la\tan ^2\delta\ra}&=&1+\left\{\begin{array}{cc}
\frac{\sqrt{1-\rho^2}}{1+\sqrt{1-\rho^2}}+ \frac{12 T}{
\Delta}\frac{XM(1-\rho^2)}{1+\sqrt{1-\rho^2}}, & T\gg \frac{N\Delta}{2\pi}\\
\frac{2\sqrt{1-\rho^2}}{4+\rho^2} +
\frac{2XM^2(1-\rho^2)^2}{4+\rho^2},
&T\ll\frac{N\Delta}{2\pi}\end{array}\right.\nonumber
\end{eqnarray}
Very important is the case of symmetric contacts, $N_1=N_2$, and
antisymmetric bias mode, $\eta=0$, which is used in
Ref.\,\onlinecite{ensslin}. Then $X$ vanishes and the average
$\tan^2\d$ becomes independent of interaction strength and as a
function of $T$ it is very weak. That is not the case if $\eta\neq
0$, for example, when only one of the voltages changes, $\eta=\pm
\pi/4$.\cite{Bouchiat,Bouchiat_preprint} Then the statistics of the
AB phase becomes temperature and interaction dependent due to the
presence of $X$.

The limit $M\gg N$ corresponds to a uniformly chaotic ring, which we
suppose to be closer to the experimental situation. Then the
dependence on $M$ drops out and the high/low temperature asymptotic
read
\begin{eqnarray}\label{eq:ANGLEwidearm}
\frac{1}{\la\tan ^2\delta\ra}&=&1+8X\left\{\begin{array}{cc}
3NT/\Delta, & T\gg N\Delta/2\pi,\\
N^2/5, &T\ll N\Delta/2\pi\end{array}\right. .
\end{eqnarray}
This result clearly demonstrates that the phase of the oscillations
is expected to deviate strongly from 0, especially if the
temperature is low and the number of channels in the contacts is
diminished. The temperature is taken into account only in the form
of temperature-averaging and the dephasing (previously considered
for nonlinear transport of noninteracting electrons in
Refs.\,\onlinecite{KY,Aronov,Yudson}) is not included.

We expect our model for chaotic AB rings to work both for
experiments at small frequencies \cite{ensslin,Bouchiat} and for
large frequencies.\cite{Bouchiat_preprint} Similarly to quantum
dots, the generalization on the finite-frequency case is obvious, if
we use Eq. (\ref{eq:Xomega}). Even in cases where RMT cannot be
assured to be valid for open diffusive rings, the dependence of the
AB phase on interaction strength, temperature, and number of
external channels given by Eq. (\ref{eq:ANGLEwidearm}) should be
correct qualitatively.

The experiment of Leturcq \etal\,\cite{ensslin} is performed in a
bias mode $\eta=0$ when $X=0$. Then Eq. (\ref{eq:ANGLEwidearm})
gives $\la\tan^2\d\ra=1$. The phase of the oscillations is evaluated
from data according to Eq. (\ref{eq:tan}) over a large range of
fields. In experiment the AB phase is varied continuously as a
function of the gate voltage at one of the arms of the ring. The
data demonstrate that the phase $\d$ indeed changes in a wide range
and is usually far from 0. This substantiates our conclusion that in
the mode when the classical contribution is minimized, $X\to 0$, the
mesoscopic distribution of $\d$ is very wide.

Experiment of Angers \etal\,\cite{Bouchiat} varies voltage in a
different way, $\eta=\pi/4$, and therefore has $X\neq 0$. We would
expect the phase $\d\mbox{(mod)}\pi$ to take values closer to $0$
and the antisymmetric component of the oscillations be relatively
smaller even for large fields. Although phases close to 0 are indeed
observed, the field averaging is taken only over first few
oscillations. In this range $\Ga$, the numerator in Eq.
(\ref{eq:tan}) is still small and grows linearly with magnetic
field. Averaging over a larger field-range similar to Ref.
\onlinecite{ensslin} could not be performed because of the phase
slips.

Another interesting question is a difference in data
\cite{ensslin,Bouchiat} for the relative magnitudes $\G_2/\G_1$ of
the second $hc/2e$ and main harmonic $hc/e$, see Eq.
(\ref{eq:expansion}). In the nonlinear transport regime this
harmonic is small compared to its contribution in the linear
transport, $\G_2/\G_1\ll G_2/G_1$,\cite{ensslin} while in
Ref.\,\onlinecite{Bouchiat} they were comparable, $\G_2/\G_1\approx
G_2/G_1$. Our model also predicts the mesoscopically averaged
contribution of $hc/2e$ into linear and nonlinear conductance to be
comparable with that of $hc/e$. Our approach assumes full quantum
coherence of the ring, and probably the difference in data is due to
decoherence.

\section{Conclusions}\label{sec:conclusions}
In this paper, we consider mesoscopic chaotic samples (quantum dots
or rings) and find the statistics of their nonlinear conductance
$\G$. This transport coefficient characterizes nonlinear DC current
due to DC-bias or a rectified current due to AC bias or
photon-assisted transport. For chaotic samples, the nonlinear effect
is of quantum origin, which is clear from the fact that its ensemble
average over similar samples vanishes. The linear response of the
sample in two-terminal measurements is always symmetric with respect
to magnetic field inversion. However, the Coulomb interactions lead
to magnetic field asymmetry of the nonlinear DC response, which
fluctuates due to the electronic interference. For the quantum dots
we consider the fluctuations of (anti) symmetrized components
$\Ga,\Gs$ of the nonlinear conductance. In chaotic rings the
statistics of the phase of AB oscillations in the nonlinear
transport regime, closely related to the ratio $\Ga/\Gs$, is of
interest.

Unlike the linear conductance measurements, in mesoscopic nonlinear
transport experiments the way voltages
 are varied ("bias mode") turns out to be important, especially for
 a weakly interacting sample. We demonstrate this fact qualitatively and
discuss the role of Coulomb interactions. Quantitative
self-consistent treatment of interactions allows us to consider
magnetic-field asymmetry in chaotic quantum dots with many channels.
Using Eqs. (\ref{eq:main})--(\ref{eq:X}) we show that the
fluctuations of $\Gs$ are strongly affected by the geometry of the
setup and discuss how the bias mode influences data of recent DC
experiments.

Another important issue is rectification of AC bias, which is
quadratic in applied voltage, random, and asymmetric with respect to
the magnetic flux inversion. The photovoltaic DC current can be due
to rectification of external perturbations or quantum pumping by
internal perturbations. Both rectification and quantum pumping share
the aforementioned properties, and it is important to clearly
separate them especially when the frequency of perturbations is high
(nonadiabatic). We consider here only the effects of the external
perturbations and discuss the dependence of the fluctuations of
$\Ga,\Gs$ on frequency $\oo$. We show that the fluctuations of both
$\Ga$ and $\Gs$, presented in Eqs.
(\ref{eq:varGs})--(\ref{eq:Xgeneral}), decrease monotonically as
$\oo\to\infty$. However, contrary to naive expectations, their
asymptotical behavior can be very different. Since at high
frequencies the response of the dot to the external bias becomes
rather capacitive then resistive, the coupling to the nearby gates
can be strongly enhanced. If the experiment is performed in a bias
mode where such coupling contributes, the symmetrized
$\Gs^2(\oo)\propto 1/\oo^3$ can become much larger then
$\Ga^2(\oo)\propto 1/\oo^6$ valid for a strongly interacting quantum
dot. The same conclusion holds in the weakly interacting limit, when
$\Gs^2\propto 1/\oo^{3/2}$ and $\Ga^2\propto 1/\oo^{3}$.

In addition, we show that recent experiments in chaotic
Aharonov-Bohm rings might be considered similarly to quantum dots.
The multiply connected geometry alone leads to AB oscillations, yet
the mesoscopic distribution of their phase is expected to be
qualitatively similar to that of $\arctan \Ga/\Gs$ in quantum dots.
Therefore, the bias mode should strongly affect the shape of
mesoscopic distribution of the AB phase. The model of an AB ring,
which we develop, consists of a dot and a long ballistic arm and
takes into account both chaos and a ring geometry. As an application
of our model we consider fluctuations of the AB phase. Unlike the AB
phase in the linear conductance, pinned to $0\mbox{(mod)}\pi$ by the
Onsager symmetry relations, the fluctuations of the AB phase in
nonlinear transport are shown to depend on the bias mode,
interaction strength, and temperature.

\section{Acknowledgements}
We thank H\'el\`ene Bouchiat, Piet Brouwer, Renaud Leturcq, David
S\'anchez, Maxim Vavilov, and Dominik Zumb\"uhl for valuable
discussions. We also thank the authors of
Ref.\,\onlinecite{Bouchiat_preprint} for sharing their results with
us before publication. This work was supported by the Swiss National
Science Foundation, the Swiss Center for Excellence MaNEP, and the
STREP project SUBTLE.

\appendix
\section{diffuson and cooperon for chaotic ring}\label{sec:appendix}
In this Appendix we determine the diffuson and cooperon
contributions to the $\S$-matrix correlators of the random
scattering matrix of a chaotic Aharonov-Bohm (AB) ring. This
calculation is performed using Random Matrix Theory (RMT).

First we explain what approximations should be made to ensure
validity of RMT. Our starting point is the assumption that the
$\S$-matrix of the ring is uniformly distributed over the unitary
group. This means that the ring is essentially zero-dimensional,
similarly to quantum dots. RMT is applicable if all energy-scales
are much smaller then the Thouless energy $\Thou$ and the total flux
through the annulus of the ring is much smaller then $\Phi_0$.
Assume the ring of radius $R$ and width $W\ll R$ to be diffusive
with diffusion coefficient $D=l v_{\rm F}/2$. To evaluate $\Thou$ we
neglect with transversal motion of an electron and find
$\Thou=\hbar/\erg=(\hbar l v_{\rm F})/2R^2$ as a solution to Laplace
equation along the circumference of the ring. RMT can be applied to
a closed ring if the dimensionless conductance is large,
$g=\Thou/\Delta=k_{\rm F}l W/2R\gg 1$, which is usually satisfied
for a weak disorder even if $W\ll R$.

\begin{figure}
  \includegraphics[width=4cm]{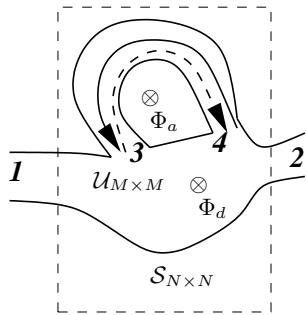}\\
  \caption{Chaotic dot combined with long ballistic multichannel arm.}\label{fig:Abdot}
\end{figure}
An open ring with ballistic contacts of the width $W_c$ gains a new
energy parameter, the escape rate $\hbar/\dwell=N\Delta/2\pi$, where
$N$ is the total number of ballistic channels. The scattering matrix
$\S$ is uniformly distributed and independent of the exact positions
of the contacts (and therefore the length of the arms) if
$\hbar/\dwell\ll\Thou\Rightarrow \pi^2 l W\gg 2R W_c$. In this case
the main drop of the potential occurs in the contacts.

If a magnetic field is applied, the RMT is valid if the total flux
through the annulus of the ring is much less then the flux quantum,
$\Phi\ll\Phi_0$. Due to narrow contacts the time-reversal symmetry
(TRS) of the $\S$-matrix can be broken at a much smaller scale,
$\Phi\sim \Phi_0\sqrt{\erg/\dwell}$. Since in our rings
$\erg\ll\dwell$, a full crossover to the broken TRS can be
considered.

How well are these conditions fulfilled in the experiment? In
Ref.\,\onlinecite{ensslin} chaos was mainly due to diffusive
scattering on the boundary and $l\approx R$. The width of the arm is
2-4 channels, while the number of channels in the contacts is $N\sim
2$, estimated from the linear conductance measurements, so
$\dwell/\erg\sim 5\div 10\gg 1$. In semiballistic samples of
Ref.\,\onlinecite{Bouchiat} (obtained by etching, and therefore
having diffusive boundary scattering) $W=W_c$ and the mean free path
is estimated $l\sim 1\div 2\mu$m$\sim L= 1.2 \mu$m, the side length.
Therefore, we have a similar estimate for the ratio $\dwell/\erg$.
Although this ratio is not parametrically large due to, e.g., weak
disorder $k_{\rm F}l\gg 1$, we believe that such AB rings still can
be assumed zero-dimensional due to their good conducting properties
together with relatively narrow contacts.

In our calculations we make a further simplification by spatially
separating chaotic scattering which randomizes the electronic phase
and the long ballistic arm attached to it. To find the correlators
of the $\S$-matrix elements we use a simplified model, see Fig.
\ref{fig:Abdot}, which combines chaos and a ring geometry. A chaotic
$M$-channel dot is attached to a long multi-channel ballistic arm
with $(M-N)/2$ orbital channels. We assume that the size of the dot
 $L$ and the length of the arm $L_a$ are such that $L_a\gg L\gg \sqrt
{(M-N)\lambda_{\rm F} L_a}$ to ensure that in the hierarchy of
different fluxes the main flux $\Phi_h$ is concentrated in the
region embraced by the arm, the flux through the dot $\Phi_d$ is
much smaller, but still much larger then the flux through the cross
section of the arm. The amplitude of AB oscillations depends on the
width of the arm $\propto (M-N)$. The wider the arm (relatively to
the contacts) the closer the results should be to a uniformly
chaotic ring. For the case when $M\gg N$ we expect it to be valid
for the chaos uniformly distributed over the ring. Indeed, in this
case an electron makes $\sim M/N\gg 1$ windings around the arm
before exiting.

In this appendix it is more convenient for us to work with an
energy-dependent matrix $\S(\e)$, and the final transformation to
time-representation is rather obvious. The total scattering matrix
$\S$ is of size $N\times N$ due to scattering channels in the
contacts 1 and 2. Chaotic scattering in the $M$-channel quantum dot
is characterized by the $M\times M$ matrix $\U$. The scattered
electron can either exit the sample through the $N=N_1+N_2$ channels
(projection operator $P_0=1_1\oplus 1_2$) or propagate into the arm
with $N_3=N_4=(M-N)/2$ channels. Electrons propagate through this
arm ballistically and gain phases which depend on the flux through
the hole. In the absence of backscattering the electronic amplitudes
at energy $\e$ are related to the path length $L_a$ and magnetic
field phase $\phi$:
\begin{eqnarray}\label{eq:Pmatrix}
\binom{b_3}{b_4}=e^{-ik(\e)L_a}\left(%
\begin{array}{cc}
  0 & e^{-i\phi} \\
  e^{i\phi}& 0 \\
\end{array}%
\right)\binom{a_3}{a_4}=P\binom{a_3}{a_4}.
\end{eqnarray}
The scattering matrix of the arm is $\P(\e)=0_1\oplus 0_2\oplus P$.
Each time an electrons enters the arm either through the third or
fourth lead, the matrix $\P$ contributes to the scattering amplitude
of the process. The total scattering matrix $\S$ is determined from
the following equation:
\begin{eqnarray}
\S=P_0\sum_{n=0}^\infty\U(\P \U)^n  P_0=P_0\U\frac{1}{1-\P\U}P_0,
\end{eqnarray}
where multiple $n\geq 0$ windings are taken into account. Both
$\U(\e,B)$ and $\P(\e,B)$ are field and energy dependent. Once we
are interested only in pair correlators of $\S(\e),\S^\dagger(\e')$
for $N,M\gg 1$, the diffuson $\Diff$ and cooperon $\Coop$ of our
scattering matrix are expressed via correlators of the dot,
$\Diff_{\cal U}$,$\Coop_{\cal U}$, and tr
$\P(\e)\P^{*(\dagger)}(\e')$. The correlators of the $\U$-matrix are
known, see Eq. (\ref{eq:cum1t}) for their time representation, and
for $\Diff$ and $\Coop$ we derive
\begin{eqnarray}\label{eq:DC}
\Brace{\Coop}{\Diff}^{-1}&=&\Brace {{\cal C}_{\cal U}^{-1}-\mbox{tr
}\P(\e)\P^*(\e')}{{\cal D}_{\cal U}^{-1}-\mbox{tr
}\P(\e)\P^\dagger(\e')},
\\
\Brace{\Coop}{\Diff}_{\cal U}^{-1}&=& M-2\pi
i\frac{\e-\e'}{\Delta}+\frac{h v_F
l}{L^2\Delta}\left(\frac{\Phi_d\pm
\Phi_d'}{2\Phi_0}\right)^2.\label{eq:DCU}
\end{eqnarray}
 The flux penetrating the dot is denoted as $\Phi_d$ and the
phase $\phi\approx 2\pi\Phi_h/\Phi_0$ gained in the arm depends on
the flux $\Phi_h$ through the hole. The traces read
\begin{eqnarray}\label{eq:tr}
\mbox{ tr
}\Brace{\P(\e,\Phi)\P^*(\e',\Phi')}{\P(\e,\Phi)\P^\dagger(\e',\Phi')}&=&(M-N)
\cos\Brace{\phi+\phi'}{\phi-\phi'}\nonumber
\\ &\times &e^{iL_a [k(\e)-k(\e')]}.
\end{eqnarray}
Since we assumed that the area of the arm is small compared to that
of the dot, the energy-dependence of Eq. (\ref{eq:tr}) can be
neglected compared to that of $\Diff_{\cal U},\Coop_{\cal U}$ in Eq.
(\ref{eq:DCU}). We also assumed that since the arm is much longer
than the size of the dot, $L_a\gg L$, the phases $\phi,\phi'$ of
open trajectories in the arm correspond to the flux $\Phi_h,\Phi_h'$
through the hole. Therefore, the effective number of channels
$N_{\Coop,\Diff}$, similar to Eq. (\ref{eq:channels}) for quantum
dots is
\begin{eqnarray}\label{eq:DCfinal}
\Brace{N_{\cal C}}{N_{\cal D}} &=& M-(M-N)\cos
\frac{2\pi(\Phi_h\pm\Phi'_h)}{\Phi_0}\nonumber \\
&&+\frac{h v_F l}{ L^2\Delta}\left(\frac{\Phi_d\pm
\Phi_d'}{2\Phi_0}\right)^2.
\end{eqnarray}
The energy-dependent cooperon and diffuson in energy representation
are given by $X(\e,\e')=1/[N_X-2\pi i(\e-\e')/\Delta],
X=\Coop,\Diff$. Notice that when $\Phi=\Phi'$ the cooperon $\Coop$
is nonperiodic in the total flux $\Phi=\Phi_h+\Phi_d$ due to finite
flux through the material of the sample, $\Phi_d\neq 0$.



\begin{thebibliography}{MMM}

\bibitem{UFN} V.~I.~Belinicher and B.~I.~Sturman, Usp.~Fiz.~Nauk {\bf 130}, 415
(1980) [Sov.~Phys.~Usp. {\bf 23}, 199 (1980)].

\bibitem{WW} S.~Washburn and R.~A.~Webb, Rep.~Prog.~Phys. {\bf 55}, 1311
(1992).

\bibitem{AK} B.~L.~Altshuler and D.~E.~Khmelnitskii, Pis'ma Zh.~Eksp.~Teor.~Fiz. {\bf 42},
291 (1985) [JETP Lett. {\bf 42}, 359 (1985)].

\bibitem{KL}D.~E.~Khmelnitskii and A.~I.~Larkin, Phys.~Scr., T
{\bf T14}, 4 (1986); Zh.~Eksp.~Teor.~Phys. {\bf 91}, 1815 (1986)
[Sov.~Phys.~JETP {\bf 64}, 1075 (1986)].

\bibitem{SB}D.~S\'anchez and M.~B\"uttiker, Phys.~Rev.~Lett. {\bf 93}, 106802
(2004).


\bibitem{SZ} B.~Spivak and A.~Zyuzin, Phys.~Rev.~Lett. {\bf 93}, 226801
(2004).


\bibitem{wei}J.~Wei, M.~Shimogawa, Z.~H.~Wang, I.~Radu, R.~Dormaier, and D.~H.~Cobden
, Phys.~Rev.~Lett. {\bf 95}, 256601 (2005).

\bibitem{marlow} C.~A.~Marlow, R.~P.~Taylor, M.~Fairbanks, I.~Shorubalko, and H.~Linke
, Phys.~Rev.~Lett. {\bf 96}, 116801 (2006).

\bibitem{ensslin}R.~Leturcq, D.~S\'anchez, G.~G\"otz, T.~Ihn, K.~Ensslin, D.~C.~Driscoll, and A.~C.~Gossard,
Phys.~Rev.~Lett. {\bf 96}, 126801 (2006).

\bibitem{Zumbuhl} D.~M.~Zumb\"uhl, C.~M.~Marcus, M.~P.~Hanson, A.~C.~Gossard,
 Phys. Rev. Lett. {\bf 96}, 206802 (2006).

\bibitem{Bouchiat} L.~Angers, E.~Zakka-Bajjani, R.~Deblock, S.~Gu\'eron, H.~Bouchiat,
 A.~Cavanna, U.~Gennser, and M.~Polianski, Phys.~Rev.~B {\bf 75}, 115309
 (2007).

\bibitem{Bouchiat_preprint} L.~Angers, A.~Chepelianskii,
R.~Deblock, B.~Reulet, and H.~ Bouchiat, Phys.~Rev.~B {\bf 76},
075331 (2007).

\bibitem{Coulomb}
D.~ S\'anchez and M.~B\"uttiker, Phys. Rev. B {\bf 72}, 201308(R)
(2005).

\bibitem{PB}M.~L.~Polianski and M.~B\"uttiker, Phys.~Rev.~ Lett.
{\bf 96}, 156804 (2006).



\bibitem{Tsvelik} A.~De~Martino, R.~Egger, and A.~M.~Tsvelik,
Phys.~Rev.~Lett. {\bf 97}, 076402 (2006).

\bibitem{PhysicaE} M. L. Polianski and M. B\"uttiker, Physica~E
(Amsterdam) {\bf 40}, 67 (2007).


\bibitem{AG} A.~V.~Andreev and L.~I.~Glazman, Phys.~Rev.~Lett. {\bf 97},
 266806 (2006).

\bibitem{Feldman}
B.~Braunecker, D.~E.~Feldman, and F.~Li, Phys.~Rev.~B {\bf 76},
085119 (2007).

\bibitem{Lofgren}A.~L\"ofgren, C.~A.~Marlow, T.~E.~Humphrey, I.~Shorubalko,
 R.~P.~Taylor, P.~Omling, R.~Newbury, P.~E.~Lindelof, and H.~Linke, Phys.~Rev.~B {\bf 73}, 235321 (2006).

\bibitem{FK}V.~I.~Falko and D.~E.~Khmelnitsky, Zh. Eksp. Teor. Fiz.
{\bf 95}, 328 (1989) [Sov. Phys. ~JETP {\bf 68}, 186 (1989)].

\bibitem{Bykov}A.~A.~Bykov, G.~M.~Gusev, Z.~D.~Kvon, D.~I.~Lubyshev,
and V.~P.~Migal, JETP~Lett. {\bf 49}, 13 (1989),

\bibitem{Liu}J.~Liu, M.~A.~Pennington, and N.~Giordano, Phys.~Rev.~B
{\bf 45}, 1267 (1992).

\bibitem{Lin} J.~J.~Lin, R.~E.~Bartolo, and N.~Giordano, Phys.~Rev.~B {\bf 45},
14231 (1992).

\bibitem{Bartolo} R.~E.~Bartolo, N.~Giordano, X.~Huang, and G.~H.~Bernstein, Phys.~Rev.~B {\bf 55},
2384 (1997).

\bibitem{BykovAB} A.~A.~Bykov, Z.~D.~Kvon, L.~V.~Litvin, Yu.~V.~Nastaushev, V.~G.~Mansurov,
 V.~P.~Migal, and S.~P.~Moschenko, Pis'ma Zh.~Eksp.~Teor.~Fiz. {\bf 58}, 538
(1993) [JETP~Lett. {\bf 58}, 543 (1993)]; A.~A.~Bykov, L.
~V.~Litvin, N.~T.~Moshegov, and A.~I.~Toropov, Superlattices
Microstruct. {\bf 23}, 1285 (1998).

\bibitem{pump} P. W. Brouwer, Phys.~Rev.~B {\bf 58}, R10135 (1998).

\bibitem{SAA} T. A. Shutenko, I. L. Aleiner, and B. L. Altshuler
Phys.~Rev.~B {\bf 61}, 10366 (2000).


\bibitem{pedersen}  M.~H.~Pedersen and M.~B\"uttiker, Phys.~Rev.~B
{\bf 58}, 12993 (1998).

\bibitem{pump_rectif} P.~W.~Brouwer, Phys.~Rev.~B {\bf 63}, 121303(R) (2001).

\bibitem{VAA} M. G. Vavilov, V. Ambegaokar, and I. L. Aleiner, Phys.~Rev.~B
{\bf 63}, 195313 (2001).

\bibitem{Kvon} J.~Q.~Zhang, S.~Vitkalov, Z.~D.~Kvon, J.~C.~Portal, and A.~Wieck,
 Phys.~Rev.~Lett. {\bf 97}, 226807 (2006).


\bibitem{DiCarlo} L.~DiCarlo, C.~M.~Marcus, and J.~S.~Harris, Phys.~Rev.~Lett. {\bf 91}, 246804 (2003).

\bibitem{Switkes} M. Switkes, C. M. Marcus, K. Campman, and A. C. Gossard,
Science {\bf 283}, 1905 (1999).


\bibitem{Moskalets_AC} M.~Moskalets and M.~B\"uttiker, Phys.~Rev.~B
{\bf 69}, 205316 (2004); {\bf 72}, 035324 (2005).

\bibitem{Vavilov05}M.~G.~Vavilov, L.~DiCarlo, and C.~M.~Marcus,
Phys.~Rev.~B {\bf 71}, 241309(R) (2005) showed that the interference
between parasitic rectification and quantum pumping produces current
asymmetric to $\Phi\to-\Phi$ even in noninteracting dot.

\bibitem{Beenakker} C.~W.~J.~Beenakker, Rev. Mod. Phys. {\bf 69}, 731 (1997).


\bibitem{ABG} I.~L.~Aleiner, P.~W.~Brouwer, and L.~I.~Glazman,
Phys.~Rep. {\bf 358}, 309 (2002).

\bibitem{PietBeen} P.~W.~Brouwer and C.~W.~J.~Beenakker,
J.~Math.~Phys. {\bf 37}, 4904 (1996).

\bibitem{iop} M.~L.~Polianski and P.~W.~Brouwer, J.~Phys.~A
{\bf 36}, 3215 (2003).

\bibitem{buttiker1} M. B\"uttiker, A. Pr\^etre, and H. Thomas,
Phys. Rev. Lett. {\bf 70}, 4114 (1993); M. B\"uttiker, J. Phys.:
Condens. Matter {\bf 5}, 9361 (1993); M. B\"uttiker, H. Thomas, and
A. Pr\^etre, Z. Phys. B {\bf 94}, 133 (1994).

\bibitem{BLF}P.~W.~Brouwer, A~Lamacraft, and K.~Flensberg,
 Phys.~Rev.~B {\bf 72}, 075316 (2005).

\bibitem{Lofgren_2004}
A.~L\"ofgren, C.~A.~Marlow, I.~Shorubalko, R.~P.~Taylor, P.~Omling,
L.~Samuelson, and H.~Linke, Phys.~Rev.~Lett. {\bf 92}, 046803
(2004).




\bibitem{ChristenButtiker}T.~Christen and M.~B\"uttiker, Europhys. Lett. {\bf 35}, 523
(1996).

\bibitem{WignerSmith} E. P. Wigner, Phys. Rev. {\bf 98}, 145 (1955);
F. T. Smith, {\it ibid} {\bf 118}, 349 (1960).

\bibitem{BP} A number of transport problems where the Wigner-Smith
matrix plays an important role are reviewed in M.~B\"uttiker and
M.~L.~Polianski, J.~Phys. A {\bf 38}, 10559 (2005).

\bibitem{deriv} P.~W.~Brouwer, S.~A.~van Langen, K.~M.~Frahm,
M.~B\"uttiker, and C.~W.~J.~Beenakker, Phys.~Rev.~Lett. {\bf 79},
913 (1997).

 \bibitem{waves} P.~W.~Brouwer, K.~M.~Frahm, and
 C.~W.~J.~Beenakker, Waves~Random~Media {\bf 9}, 91 (1999).


\bibitem{PVB} M.~L.~Polianski, M.~G.~Vavilov, and P.~W.~Brouwer,
 Phys. Rev. B {\bf 65}, 245314 (2002).

\bibitem{PietMarkus} P.~W.~Brouwer and M.~B\"uttiker, Europhys.\ Lett. {\bf 37},
441 (1997).

\bibitem{Hekking}F.~W.~J.~Hekking and J.~P.~Pekola, Phys.~Rev.~Lett. {\bf 96},
056603 (2006).

\bibitem{KY}V.~E.~Kravtsov and V.~I.~Yudson, Phys.~Rev.~Lett.
{\bf 70}, 210 (1993).
\bibitem{Aronov} A.~G.~Aronov and V.~E.~Kravtsov, Phys.~Rev.~B {\bf 47},
 13409 (1993).
\bibitem{Yudson} V.~I.~Yudson, Phys.~Rev.~B {\bf 65}, 115309 (2002).
\end{thebibliography}
\end{document}